\documentclass[structabstract]{aa}  
\usepackage{graphicx}
\usepackage{txfonts}
\usepackage{natbib}
\newcommand{\mincir}{\raise
-2.truept\hbox{\rlap{\hbox{$\sim$}}\raise5.truept\hbox{$<$}\ }}
\newcommand{\magcir}{\raise
-2.truept\hbox{\rlap{\hbox{$\sim$}}\raise5.truept\hbox{$>$}\ }}

\begin{document}
   \title{Angular correlation functions of X-ray point-like sources in
     the full exposure XMM-LSS field {\bf \thanks{This paper is
         dedicated to the memory of Olivier Garcet who has initiated the present work just before his sudden death.}} }%and Subaru
                                %fields}
%   \subtitle{...} 

   \author{A.~Elyiv\inst{1,}\inst{2} \and 
 N.~Clerc\inst{3} \and
 M.~Plionis\inst{4,}\inst{5} \and
 J.~Surdej\inst{1} \and
 M.~Pierre\inst{3} \and
 S.~Basilakos\inst{6,}\inst{7} \and
L.~Chiappetti\inst{8} \and
  P.~Gandhi\inst{9} \and
 E.~Gosset\inst{1} \and
 O.~Melnyk\inst{1,}\inst{10} \and  
F.~Pacaud\inst{11}}

   \offprints{A. Elyiv}

\institute {Institut d'Astrophysique et de G\'eophysique, Universit\'e
  de Li\`ege, 4000 Li\`ege, Belgium
 \and Main Astronomical Observatory, Academy of Sciences of Ukraine,
 27 Akademika Zabolotnoho St., 03680 Kyiv, Ukraine
  \and DSM/Irfu/SAp, CEA/Saclay, F-91191 Gif-sur-Yvette Cedex, France
 \and Institute of Astronomy \& Astrophysics, National Observatory of
 Athens, Thessio 11810, Athens, Greece
 \and Instituto Nacional de Astrof\'isica, \'Optica y Electr\'onica, 72000 Puebla, Mexico
 \and Academy of Athens, Research Center for Astronomy and Applied
 Mathematics, Soranou Efesiou 4, 11527, Athens, Greece
\and High Energy Physics Group, Dept. ECM, Universitat de
  Barcelona, Av. Diagonal 647, E-08028 Barcelona, Spain
 \and INAF-IASF Milano, via Bassini 15, I-20133 Milano, Italy
 \and Institute of Space and Astronautical Science (ISAS), Japan
 Aerospace Exploration Agency, 3-1-1 Yoshinodai, Chuo-ku, Sagamihara,
 Kanagawa 252-5210, Japan
 \and Astronomical Observatory, Kyiv National University, 3 Observatorna St., 04053 Kyiv, Ukraine
 \and Argelander Institut f\"ur Astronomie, Universit\"at Bonn, Germany}

\date{Received 31 August 2011 / Accepted 13 November 2011}

 % \abstract{}{}{}{}{} 
% 5 {} token are mandatory
 
  \abstract
   {}
   {Our aim is to study the large-scale structure of different types
     of AGN using the medium-deep XMM-LSS survey.}
{We measure the two-point angular correlation function of $\sim
  5700$ and 2500 X-ray point-like sources over the $\sim 11$ sq. deg. XMM-LSS field
  in the soft (0.5-2 keV) and hard (2-10 keV) bands. For the
  conversion from the angular to the spatial correlation function we 
  used the Limber integral equation and the luminosity-dependent density
evolution model of the AGN X-ray luminosity function.}
   {We have found significant angular correlations %in the case of
     %the whole XMM-LSS field 
     with the power-law parameters 
     $\gamma =1.81\pm 0.02$, $\theta _{0}=1.3''\pm
     0.2''$ for the soft, and  $\gamma =2.00\pm 0.04$, $\theta
     _{0}=7.3''\pm 1.0''$ for the hard bands. The amplitude of the
     correlation function $w(\theta)$ is higher in the hard than in
     the soft band for $f_x\mincir 10^{-14}$ erg s$^{-1}$ cm$^{-2}$ and lower
     above this flux limit. We confirm that the
     clustering strength $\theta _{0}$ grows with the flux limit of
     the sample, a trend which is also present in the
     amplitude of the spatial correlation function, but only for the
     soft band. In the hard band, it remains almost constant with $r_0 \simeq 10
     \;h^{-1}$ Mpc, irrespective of the flux limit. Our
     analysis of AGN subsamples with different hardness ratios
     shows that the sources with a hard-spectrum are more 
     clustered than soft-spectrum ones. This result may be a hint that the two main types of AGN
     populate different environments. Finally, we find that our
     clustering results correspond to an X-ray selected AGN bias
     factor of $\sim 2.5$ for the soft band sources (at a median $\bar{z}\simeq 1.1$) 
     and $\sim 3.3$ for the hard band sources 
     (at a median $\bar{z}\simeq 1$), which translates into a
     host dark matter halo mass of $\sim 10^{13} \; h^{-1} M_{\odot}$ and
     $\sim 10^{13.7}\; h^{-1} M_{\odot}$ for the soft and hard bands, respectively.

}
{}


   \maketitle
%
%________________________________________________________________

\section{Introduction}

The study of the large-scale structure for the universe and of
structure formation processes makes it necessary to carry out 
wide-field surveys of extragalactic objects. These surveys are performed
in almost all accessible wavelength bands. X-ray surveys constitute
an important part of these surveys because of the weak absorption at such high
energies. The most recent and prominent observational X-ray results have
been obtained with the XMM-Newton and Chandra space observatories 
\citep{Brandt}. More than 95{\%} of all detected objects in X-ray
surveys away from the galactic plane are point-like and 
predominantly active galactic nuclei (AGN), the rest are
mostly extended sources (groups and clusters of galaxies and
relatively nearby galaxies).
Owing to their high X-ray
luminosity, AGN can be detected over a wide
range of redshifts in contrast to normal galaxies \citep{Hartwick}, and therefore these objects
are excellent tracers of the cosmic web and a convenient tool for studying 
evolutionary phenomena in the Universe. 
It is known that the optical and
X-ray classification of type 2 (obscured) AGN agree quite well, see for
example \citet{Garcet} and references therein.
X-ray selected AGN also provide a relatively unbiased census of the 
AGN phenomenon because obscured AGN, which are largely missed in optical surveys, 
are included in X-ray surveys.

The clustering pattern of the AGN population can provide important
information regarding the cosmography of
matter density fluctuations at different scales and the cosmological
parameters \citep[e.g.,][]{Hickox07, Engels, Plionis10, ebrero,
  basilakos09, Basilakos10}, the evolution of the AGN phenomenon
\citep[e.g.,][]{Comastri, Koulouridis, Allevato}, 
the relation between AGN activity and their dark matter halo hosts, supermassive black hole formation \citep[e.g.,][]{Mandelbaum2009, Miy11, 
  Allevato}, and so on. The most common approach to quantify AGN
clustering, without redshift information 
is to measure the AGN two-point angular correlation function
\citep[ACF;][]{
%Ghaffarsedeh,
akylas_georgantopoulos_2000, Yang_2003, Manners_2003, basilakos_2005, 
gandhi06, Puccetti, Miyaji_2007, Carrera, Garcet, ebrero}, 
which provides an estimate of how significant the excess of AGN
pairs is, within some projected angular separation over that of a random
distribution. Once the angular correlation function is measured, it 
is possible to reconstruct the spatial clustering, under some specific assumptions,
 using the Limber integral equation \citep{Limber, peebles}. However,
spectroscopic follow-up as well
as multiwavelength photometric observations in a number of different bands
allow us to measure or estimate redshifts for a large number of AGN and to apply
the direct spatial correlation analysis \citep[e.g.,][]{Gilli_2005, yang06, 
gilli, Coil, Cappelluti, Miy11}. 

Clustering analyses of the various surveys of X-ray selected AGN in the soft and
hard bands have provided a wide range of angular and spatial 
clustering lengths. Strong indications for a flux-limit clustering dependence appear to reconcile most of the diverse results, however 
\citep{Plionis08, ebrero, Krumpe}.

Another important question is whether the clustering of X-ray selected
AGN evolves with time. \cite{gilli} did not find any significant 
difference between the X-ray AGN clustering below and above
$z=1$. Even so, the X-ray AGN bias factor should evolve with
time, and indeed \cite{yang06} found a rapid increase of the bias factor
with redshift with $b(z=0.45)=0.95\pm 0.15$ and $b(z=2.07)=3.03\pm
0.83$. Similarly, \cite{Allevato}  estimated the average bias in the
COSMOS AGN survey and found a redshift evolution of the bias factor
with $b(z=0.92)=2.30\pm 0.11$ and $b(z=1.94)=4.37\pm 0.27$. 

The AGN clustering pattern can also be used for the verification of
the unification model, because both obscured and unobscured AGN
should have identical correlation function, if the orientation of the
torus is the only determining factor of the AGN phenomenology. 
%Various criteria can be used for splitting into these two
%subsamples. 
\cite{gilli} used the 2 sq. deg. XMM-COSMOS field and did not find any
significant difference in the spatial distribution of the broad and
narrow line AGN. Similarly, \cite{ebrero}, studying 1063 XMM-Newton
observations, found consistent correlation properties for sources with
high and low hardness ratios, which mostly correspond to obscured and
unobscured AGN, respectively. These results postulate that obscured and
unobscured objects populate similar environments, which agrees with the unified model of AGN.

However, the analysis of the 9 sq. deg. Bootes multiwavelength survey
showed slightly different clustering properties for the two types of AGN
 \citep{hickox11}.
% log($M_{halo}$ [$h^{-1}$ $M_{sun}$]) = $12.7^{+0.4}_{-0.6}$ and
% $13.3^{+0.3}_{-0.4}$, respectively. 
Similarly, \cite{Puccetti} investigated the central 0.6
sq. deg. region of the ELAIS-S1 field and found that the
correlation amplitude in the hard band ($\theta_{0}=12.8''\pm 7.8''$) is 2.5
times higher than that in the soft band ($5.2''\pm 3.8''$), but with a weak
significance ($\sim 1\sigma$). \cite{gandhi06} used the hardness
ratio ($HR$) and divided the point-like sources in mainly obscured
($HR>-0.2$) and unobscured ($HR<-0.2$) subsamples, finding a
positive clustering signal only for the obscured sources in the hard band.

In this work we will revisit these questions by presenting 
the final results of the point-like source distribution 
of the XMM-Newton Large Scale Structure (XMM-LSS) survey of
\cite{pierre04}. In \cite{gandhi06} we presented the AGN clustering
results based on the previous release of 4.2 sq. deg. of this
survey.  A weak positive correlation signal was found 
in the soft band (angular scale $\theta_{0}=6.3''\pm
3''$ with a slope $\gamma =2.2\pm 0.2$).  At present the full XMM-LSS 
field is one of the widest ($\sim$11 sq. deg) medium-deep surveys. It is
part of an even larger project, the XXL, observations of which are
currently being implemented \citep{pierre11}.

In the following sections we present the description of the XMM-LSS survey
(Section 2). Results of the numerical simulations of the X-ray
point-like sources are presented in Section 3. Basic properties
of the XMM-LSS field, like its source distribution 
on the sky and the point-source $\log N-\log S$ relation, are shown in Section 4. 
Section 5 includes the method used to produce the random
catalogs and the ACF analysis for the different samples. 
Inverting from angular to spatial clustering and the derived bias
of AGN are presented in Sections 6 and 7, respectively, while the main conclusions
are listed in Section 8.

\section{The sample of X-ray point-like sources}

In the present correlation function analysis, we have used point-like
X-ray sources from the 
%full exposure 
XMM-LSS field, which consists of
 %and Subaru fields. The XMM-LSS field is represented by 
87 pointings with maximum available exposures
from 10 to 28 ks. Also we used 7 pointings of the independent deeper Subaru/XMM-Newton Deep Survey (SXDS)  \citep{Ueda}  whose data we reanalyzed with our
     pipeline because it is fully enclosed in the XMM-LSS area, although with a different spacing pattern. For S01 pointing of SXDS  we kept only 40 ks chunk to prevent possible source confusion.

%This field consists of 7 pointings with exposures from 19 to 49 ksec.
 Altogether, the
%full exposure 
XMM-LSS field is contiguous and contains $\sim 5700$ %$\sim 5500$ point-like 
sources in the soft (0.5 - 2 keV) band and $\sim 2500$ in the hard (2 - 10 keV)
band, out of which $\sim 180$ are extended (mainly galaxy clusters).
Although all extended sources were removed from our analysis, it is
interesting to note that they were categorized according to their
extension likelihood\footnote{By the term "likelihood" as used hereinafter, we mean  formally the log-likelihood.} (EXTlike) and core radius (EXT) into two classes:
"C1" which are the true
extended sources with EXTlike$~>~$33 and EXT$~>~5''$, containing 54 objects,
and "C2" which is a class with
almost 50 per cent contamination at 15$~<~$EXTlike$~<~$33 and EXT$~>~5''$,
containing 129 objects (see Pacaud et al. 2006 for details).

Separately, we have made use of a more uniform XMM-LSS survey for our analysis which consists of 10 ks chunks.
This catalog will be published soon (Chiappetti et al., in prep.)

We considered all point-like sources as AGN, although we do expect a $\sim 3\%$ stellar contamination \citep{Salvato}. More details about the source classification will be given in Melnyk et al. {\em in prep}.
The sensitivity limits of the joined sample are near $10^{-15}$
and $3\times 10^{-15}$ erg s$^{-1}$ cm$^{-2}$ for the soft and hard
bands, respectively.

\begin{figure}
\includegraphics[width=\columnwidth]{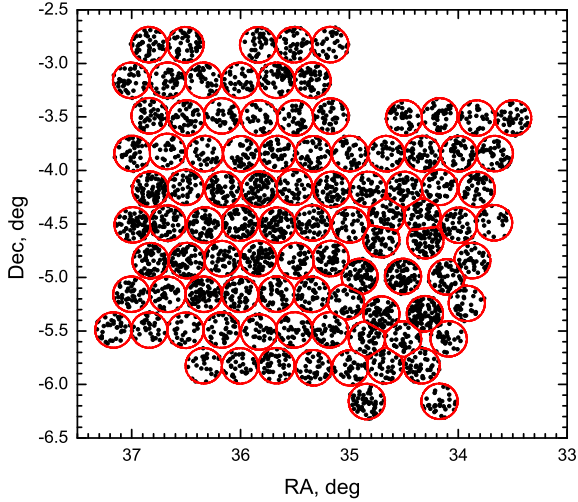}
\caption{Distribution of the X-ray point-like sources observed in the
  soft band within the whole XMM-LSS field %and Subaru fields 
  with an off-axis
  distance less than $10'$. The red circles represent the borders between
  the different pointings. Note that even when using the $10'$
  limitation, we may have some overlapping regions.%, especially near
  %the border between the Subaru and the XMM-LSS fields. 
  We discarded these minor overlaps using the Voronoi tessellation method.}

\label{fig:map10}
\end{figure}

\begin{figure} 
\includegraphics[width=\columnwidth]{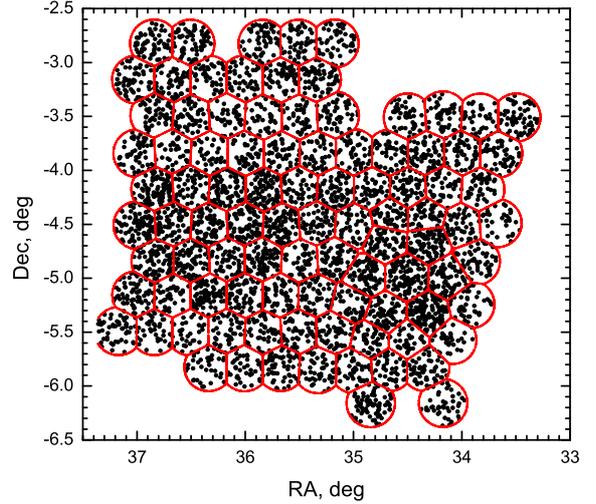}
\caption{Distribution of the X-ray point-like sources observed in the
  soft band within the whole XMM-LSS field %and Subaru fields. The red
                                %contours represent the borders in
                                %accordance
  with the Voronoi tessellation method. Note that using the
    Voronoi tessellation we did not exceed the $13'$
    off-axis distance for any pointing.}
\label{fig:mapvor}
\end{figure}

The average distances between the centers of adjacent pointings were
substantially shorter than the FoV diameter of the EPIC cameras 
to gain a more homogeneous coverage. This caused overlaps between adjacent pointings. 
Because each pointing was processed individually, the final merged
catalog was produced a posteriori (Pierre et al. 2007; Chiappetti et al., in prep). As a first
possibility, we only considered sources
with an off-axis distance $<~10'$. In this way, we did not have to
consider boundary sources that are often detected with large errors. The total
effective area of the fields was 8.3 sq. deg. The
distribution of the corresponding 4066 X-ray sources located within the borders 
is shown in Fig.~\ref{fig:map10}. As an alternative merging
possibility we applied a Voronoi boundary
delimitation \citep{matsuda} without imposing an off-axis
cutoff and using sources across the full exposure XMM-LSS field.
In other words, in any overlap region among two pointings
we only kept those detections in our final catalog that had the shortest
off-axis distance. In this way, we were able to use the widest possible
area 10.9 sq. deg. of the investigated field. The distribution of the
5093 point-like X-ray sources is shown in
Fig.~\ref{fig:mapvor}. The basic ACF results were
checked considering both approaches and it was found that apart from
larger uncertainties in the case of the $<10'$ delimitation method,
the results were statistically identical.

\section{Simulations of XMM-LSS AGN}
For a proper correlation function analysis we need to know the values
of the detection probability for each registered source. Therefore we
performed extensive simulations of each individual XMM
pointing. We briefly describe our procedure and the set of simulations
that were used for the representation of the XMM-LSS field.

	\subsection{Description of the XMM-Newton point-source simulation}
The principle of the simulations is similar to the one presented in
\cite{Pacaud:2006p3257} and \cite{gandhi06}. The main steps of the procedure consist in
i) generating an input source list drawn from a fiducial flux
distribution that is randomly distributed across the pointing field of
view;
ii) simulating images of the field as it would be seen by XMM-Newton
by reproducing the main instrumental effects (vignetting, PSF
distortion, detector masks, background and Poisson noise);
iii) detecting sources with the XMM-LSS pipeline
\citep{Pacaud:2006p3257} and obtaining their likelihoods and measured
count-rates\footnote{Throughout this paper, count-rates are expressed
  in terms of total MOS1+MOS2+PN count-rates, corrected for
  vignetting. That is why two sources with the same count-rate but different
  off-axis positions will have different probabilities of detection.};
iv) correlating the detected source list with the input catalog using
a $6''$ radius and deriving the rates of true and false detections as
well as the detection probabilities.

The simulations were performed in the soft and the hard bands.
The original source distribution was taken from
\cite{Moretti:2003p1679} using either their soft or hard band fitting
formulae, down to a flux which approximately corresponds to 2 photons
on-axis (i.e. below XMM-Newton detection limit). This value depends on
the exposure time chosen for each particular simulation.
Non-resolved AGN photon background was added following
\cite{Read:2003p3205}, then we subtracted the contribution of the AGN
resolved by our detection algorithm.  The constant conversion factor
$cf$ between the total count-rates and the physical fluxes $S$ was
calculated on the basis of the MOS and PN camera factors provided by
\cite{pierre07}.
% $cf=9.375\times 10^{-13}$ for the soft and $4.325\times 10^{-12}$ erg cm$^{-2}$ cts$^{-1}$ for the hard band. 

Particle background was also added according to values quoted in
\cite{Read:2003p3205} and was subsequently modified by multiplying these
values by an arbitrary factor between $0.1$ and $8$ to allow
for pointing-to-pointing background variations. In any case, this
component was not vignetted. We summarize our typical background values in Table \ref{table_background}. The PSF model was
taken from the XMM-Newton medium model calibration files. The
vignetting was modeled through its off-axis variation onto each
detector.

The detection algorithm provides for each source an estimate of its
count-rate on each detector as well as the local background value at
the source position. A key parameter is the source detection likelihood. Following \cite{Pacaud:2006p3257}, this quantity was computed
  using the C-statistic. Its value is the difference between the
  likelihood of the best-fitting point-source model and the likelihood
  of a pure background fluctuation. As such, the source likelihood LH
  represents the significance of the detection.
A value of $15$ provides a good balance between contamination and
completeness (see Pacaud et al. 2006 and paragraph
\ref{simulation_set} for a discussion of the stability of this
criterion).

\begin{table}
       \centering
               \begin{tabular}{@{}c|rrrr@{}} \hline & \multicolumn{2}{c}{Photon background} & \multicolumn{2}{c}{Particle background}\\ & soft band & hard band & soft band & hard band  \\ \hline
MOS1  & 1.21&1.77 &  0.764&1.16  \\
MOS2  & 1.32&1.88 &  0.730&1.09  \\
PN    & 2.49&3.55 &  2.80&6.03  \\
\hline
               \end{tabular}
               \caption{\label{table_background}Typical background
                  values for the pointing simulations. We allowed the
                  particle background to vary from one pointing to the
                  other through a multiplicative factor chosen among
                  0.1, 0.25, 0.5, 1, 2, 4 and 8. Units for each
                  XMM-Newton detector are $10^{-6}$ cts s$^{-1}$ pixel$^{-1}$.} \end{table}

	\subsection{Set of simulations}
	\label{simulation_set}
	
\begin{figure*}
	\begin{tabular}{ccc}
		\includegraphics[height=57mm]{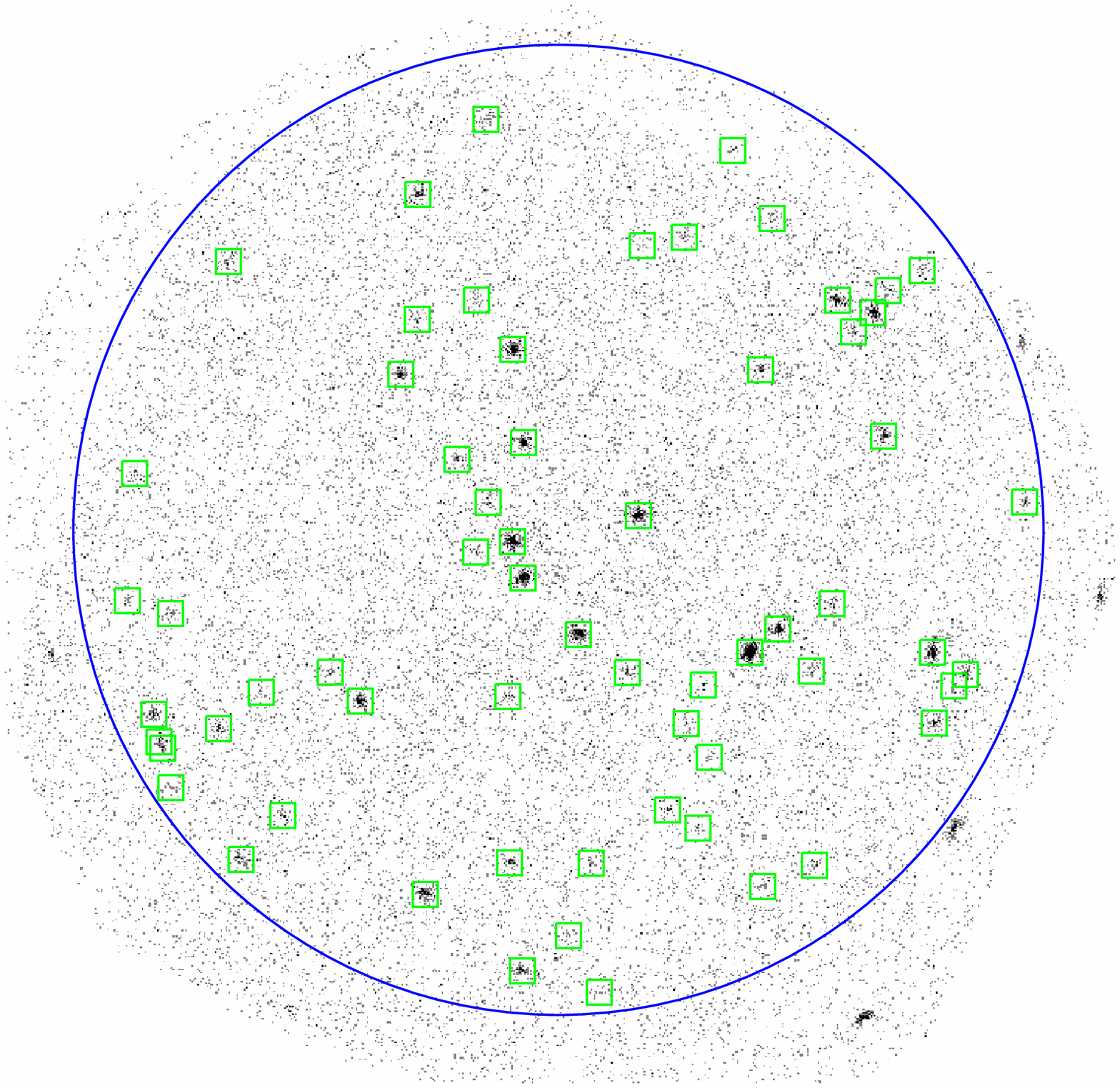} &
		\includegraphics[height=57mm]{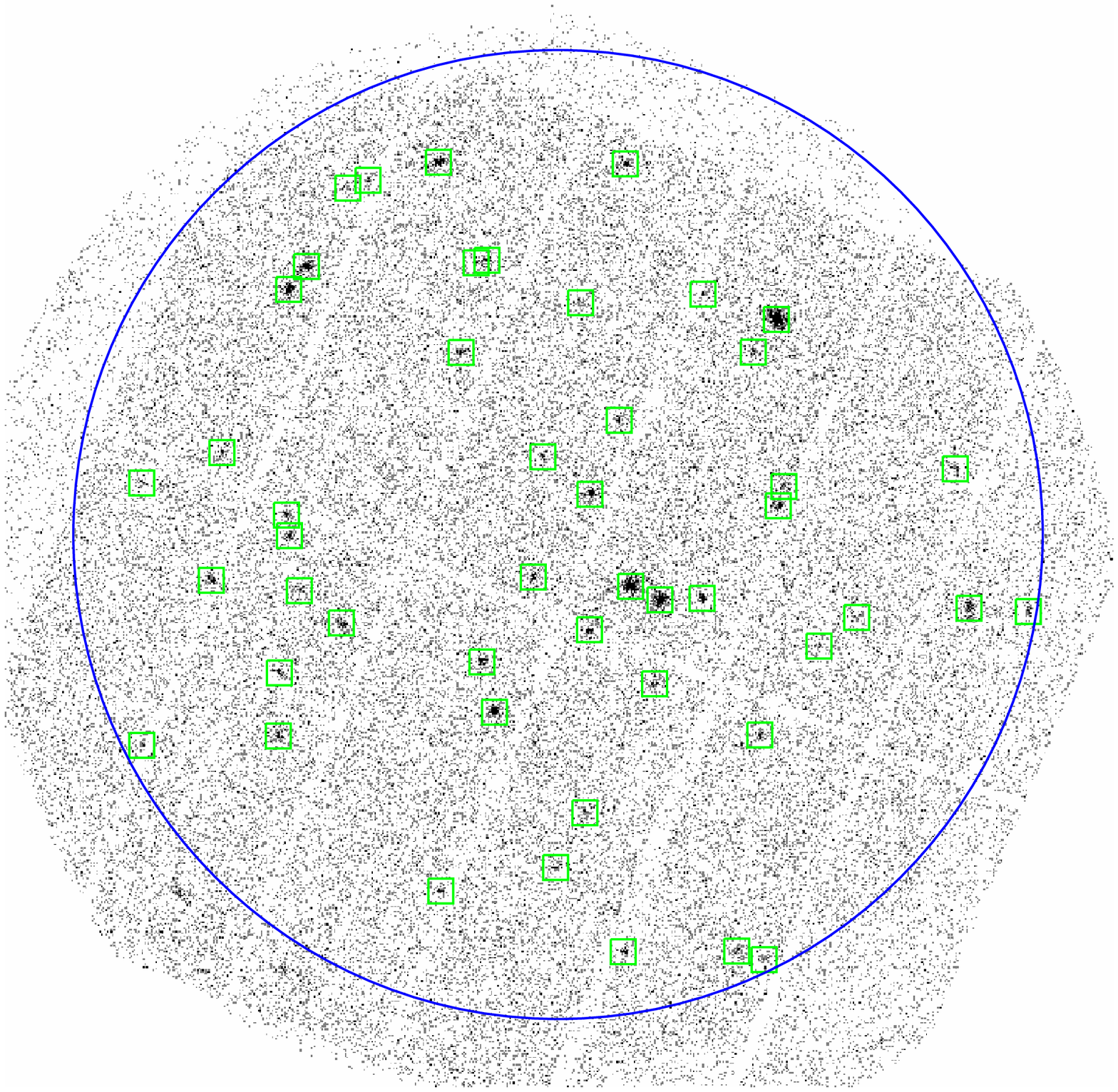} &
		\includegraphics[height=57mm]{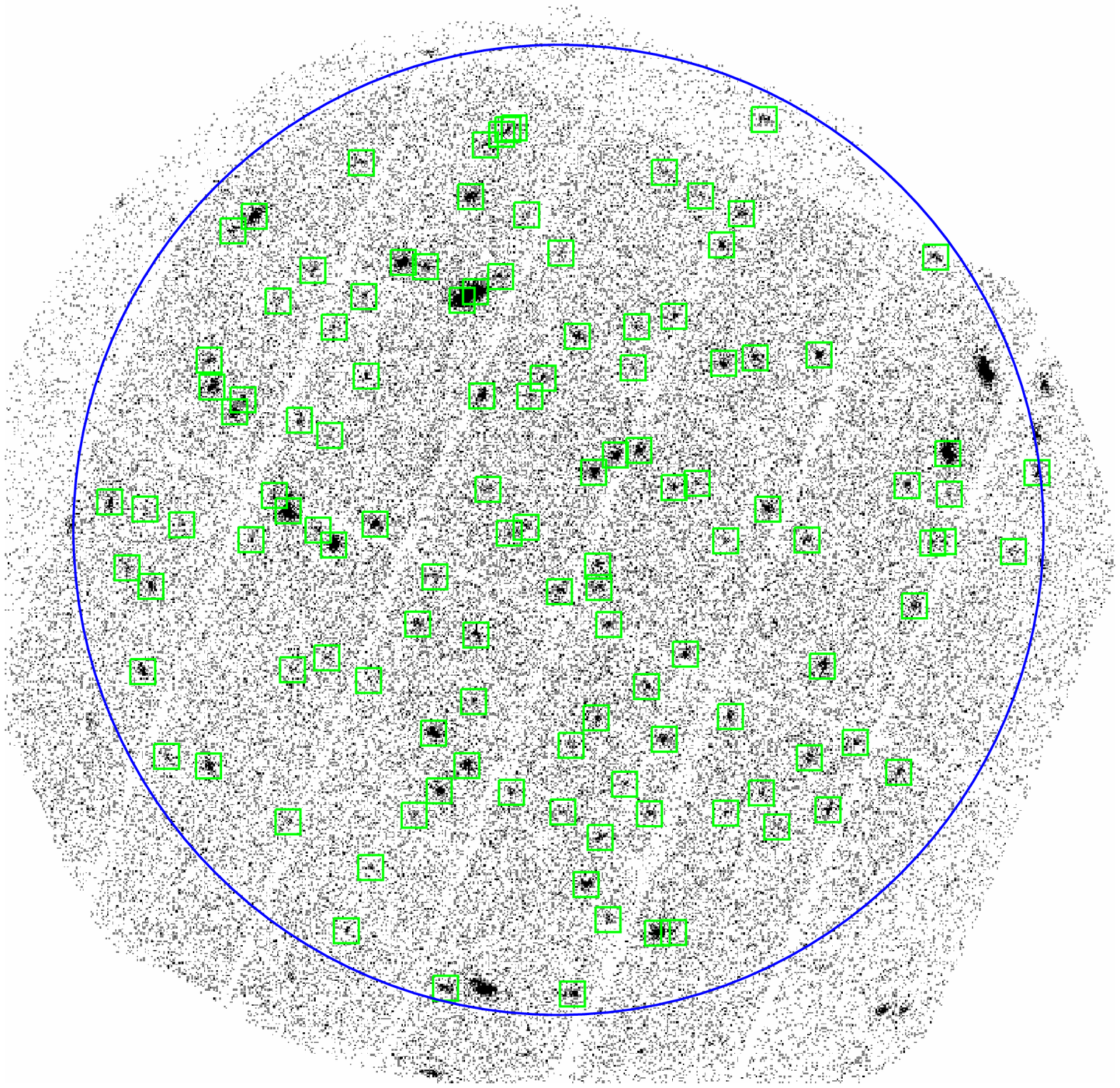} \\
	\end{tabular}
 \caption{Examples of simulated XMM-Newton pointings in the soft band.
 The large blue circle indicates the region in which the source
 detection is performed ($13$ arcmin maximal off-axis angle).
 Green boxes indicate the positions of sources detected with a likelihood ML$~>15$.
 {\it Left: } $T_{\mathrm{exp}}=10$ ks, background ratio $=1$; 
 {\it Middle: } $T_{\mathrm{exp}}=10$ ks, background ratio $=4$; 
 {\it Right: } $T_{\mathrm{exp}}=40$ ks, background ratio $=1$.
 }
 \label{fig_simus_point} 
\end{figure*}

To fully account for the variations of the detection efficiency across
the XMM-LSS fields, we simulated 18900 and 6480 pointings in the soft
and the hard bands, respectively. Table \ref{table_simset} details the
simulation set. Fig. \ref{fig_simus_point} illustrates the influence
of exposure time and background ratio value for three pointings from
our simulation set at 10 and 40 ks and for background ratios 1 and 4.

\begin{table}
	\centering
		\begin{tabular}{@{}c|ccc@{}}
\hline
 $T_{\mathrm{exp}}$	&	Limiting flux	&	Particle background	&	Number\\
 		(ks)	&	($10^{-16}$ erg s$^{-1}$ cm$^{-2}$)	&	factors		&	of fields\\
\hline
\multicolumn{2}{l}{Soft band $0.5-2$ keV}	&	&		\\
\hline
7 &	$1.43$		&	0.1 0.25 0.5 1 2 4 8		&	7 $\times$ 540		\\
10 &	$1$		&	0.1 0.25 0.5 1 2 4 8		&	7 $\times$ 540		\\
20 &	$0.5$		&	0.1 0.25 0.5 1 2 4 8		&	7 $\times$ 540		\\
40 &	$0.25$		&	0.1 0.25 0.5 1 2 4 8		&	7 $\times$ 540		\\
80 &	$0.125$	&	0.1 0.25 0.5 1 2 4 8		&	7 $\times$ 540		\\	
\hline
\multicolumn{2}{l}{Hard band $2-10$ keV}												&								&		\\
\hline
7  &	$14.3$		&	0.1 1 3		&	3 $\times$ 540		\\
10 &	$10$		&	0.1 1 3		&	3 $\times$ 540		\\
20 &	$5$		&	0.1 1 3		&	3 $\times$ 540		\\
40 &	$2.5$		&	0.1 1 3		&	3 $\times$ 540		\\
\hline
		\end{tabular}
		\caption{\label{table_simset} Summary of the soft and
                  hard band simulation sets. The second column refers
                  to the lowest flux of the input simulated sources in
                  the band of interest. Different background levels
                  are accounted for by applying a multiplicative
                  factor to the values from Table
                  \ref{table_background}. This set of simulation
                  encompasses most of the XMM-LSS pointing
                  characteristics.}
\end{table}

Thanks to the high number of simulated fields, we were able to bin our
results by source off-axis angle. We have chosen six annuli of equal
area to obtain approximately the same level of significance in each
bin. Values defining the bin bounds are 0, 5.3, 7.5, 9.2, 10.6,
11.9 and 13.0 arcmin.

\subsubsection{Completeness/contamination balance}
		
\begin{figure*}
	\begin{tabular}{ccc}
		\includegraphics[height=57mm]{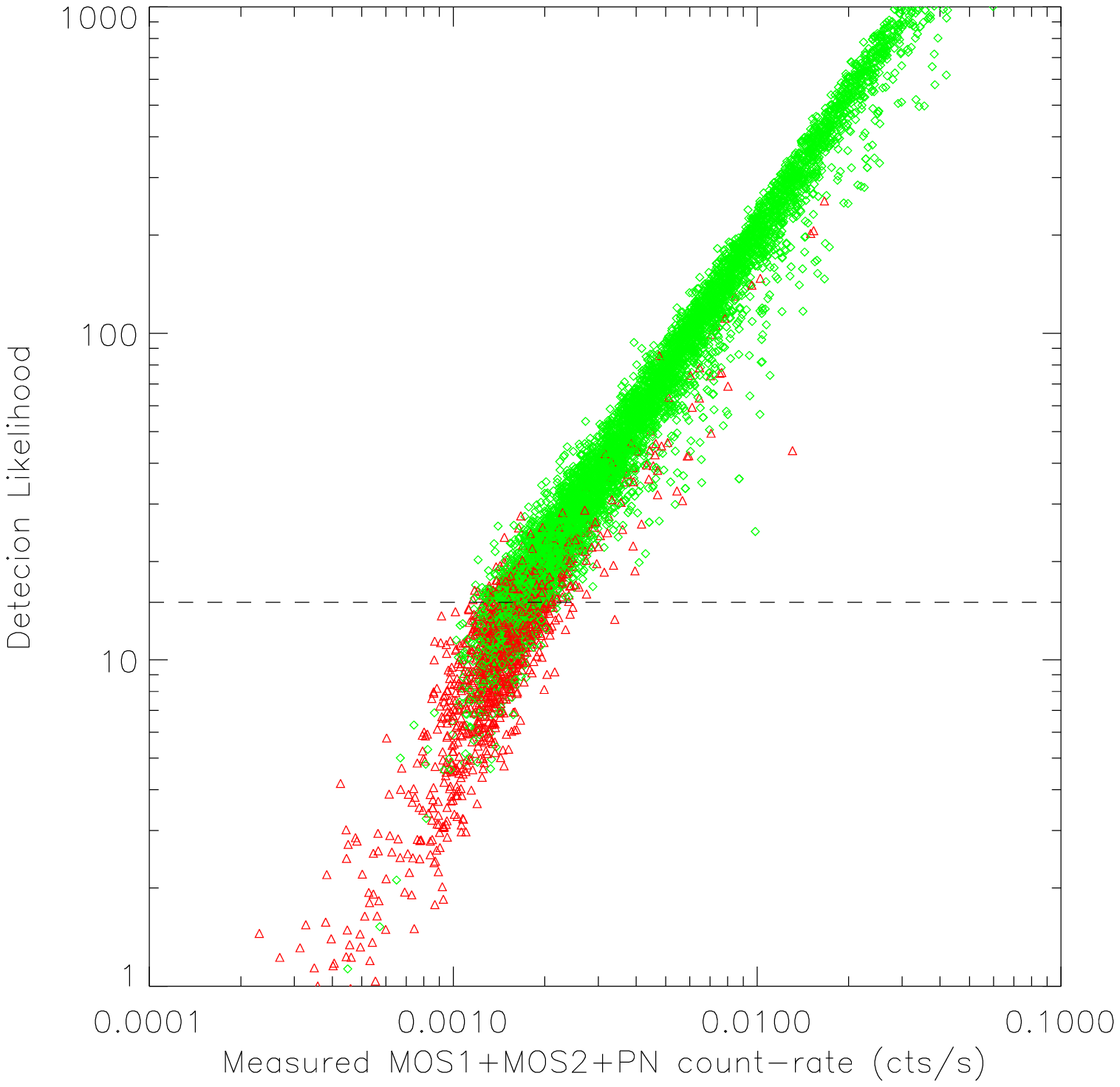}          &
		\includegraphics[height=57mm]{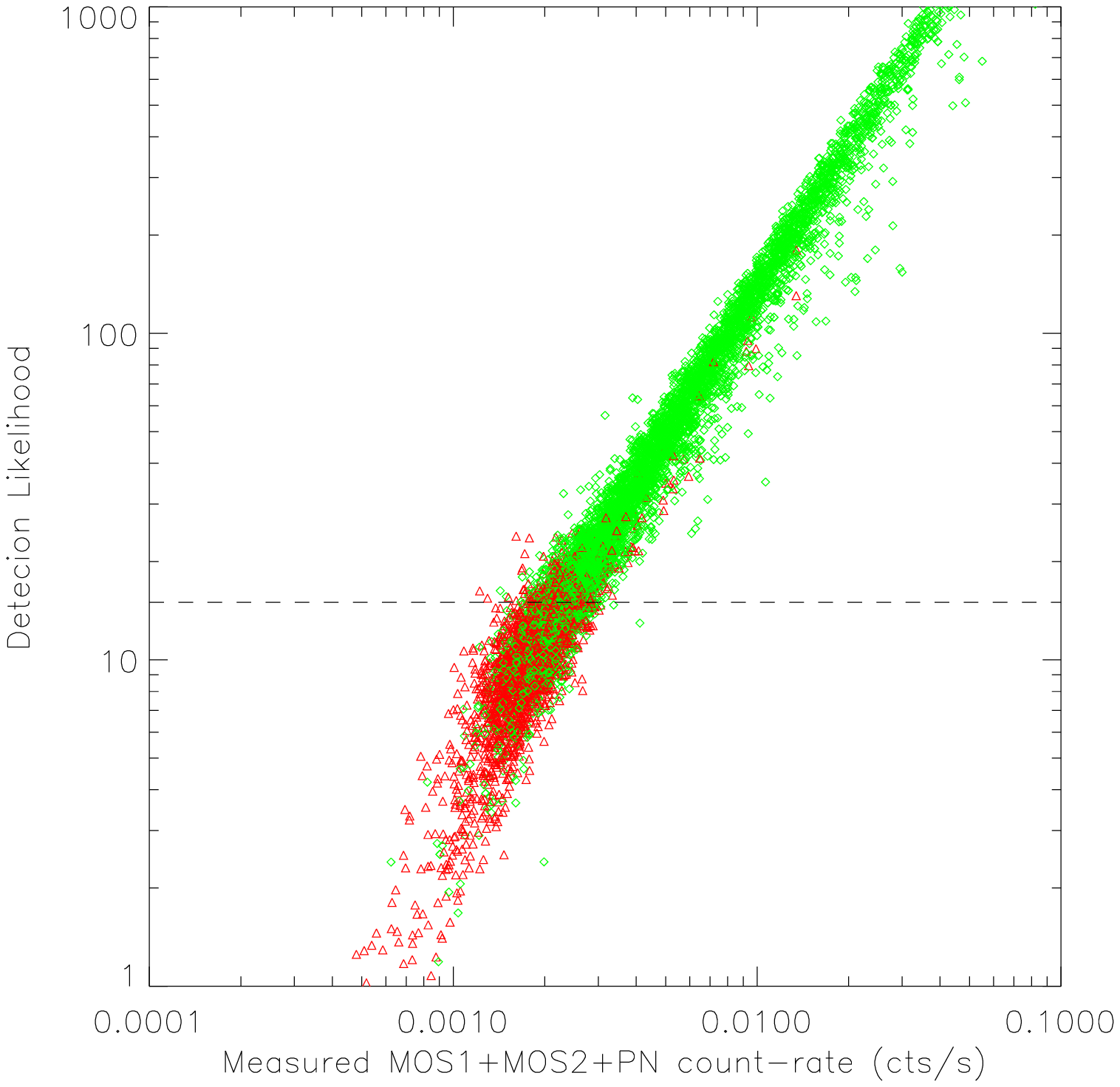} &
		\includegraphics[height=57mm]{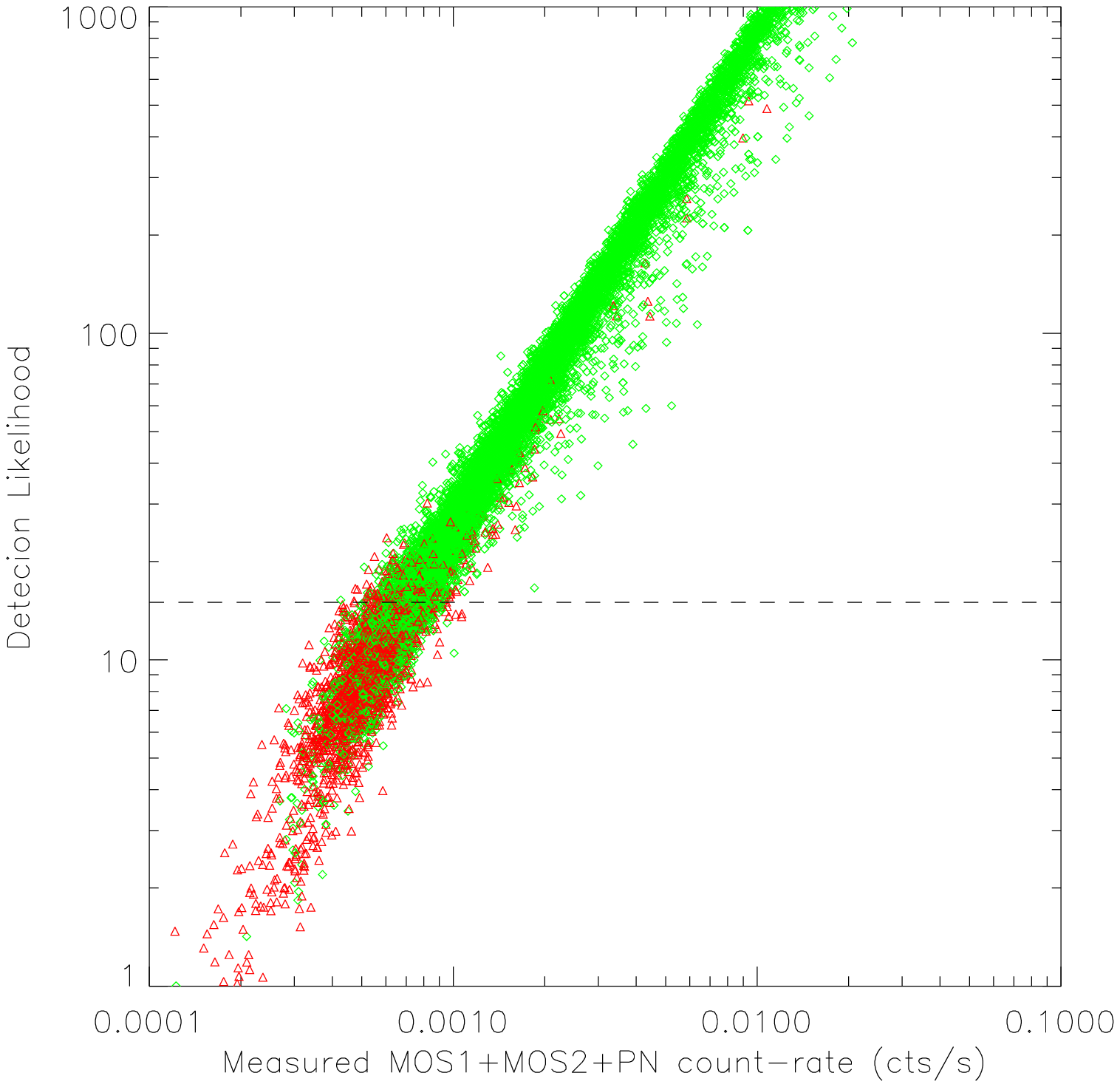} \\
	\end{tabular}
 \caption{Examples of likelihood versus count-rate plots for three 
combinations of pointing exposures and background ratios
   % pointing configurations 
from our soft band simulation set.
 Green symbols show detections with a real input counterpart while
 red points indicate spurious (false) detections.
 All sources within 5 arcmin from the pointing center are shown here.
 {\it Left: } 540 pointings with $T_{\mathrm{exp}}=10$ ks, background ratio $=1$;
 {\it Middle: } 540 pointings with $T_{\mathrm{exp}}=10$ ks, background ratio $=4$; 
 {\it Right: } 540 pointings with $T_{\mathrm{exp}}=40$ ks, background ratio $=1$.
 The horizontal line corresponds to ML$=15$, the threshold above which
 detected sources are included in the catalog.
 The separation between false and real detections is relatively
 independent of the pointing quality.
 }
 \label{fig_countdetml} 
\end{figure*}

The source selection was based on the detection likelihood value (ML);
all detected sources with ML$>15$ were included in the final sample
of point-like sources. A fraction of these sources comes from
false detections. Fig. \ref{fig_countdetml} shows the distribution of
the recovered sources for three configurations in the innermost
off-axis bin ($0-5.3$ arcmin). We see little dependence of the
contamination rate on the background level and exposure time, and the
ML$=15$ threshold appears as the best choice for homogeneous
balance between completeness of the sample and contamination by
spurious sources.

We numerically computed the rate of false detections as a function of the
off-axis angle, background level and exposure time and in any
configuration. The average rate of spurious detections is between 2
and 5 per pointing (up to $13$ arcmin off-axis angle). A typical
pointing ($T_{\mathrm{exp}}=10$ ks, background ratio $=1$) gives from
40 to 50 detections in the $[0-10]$ arcmin off-axis, which leads to a
contamination rate of about 5\%.

\subsubsection{Detection efficiency as a function of pointing characteristics}

\begin{figure}
	\includegraphics[width=91mm]{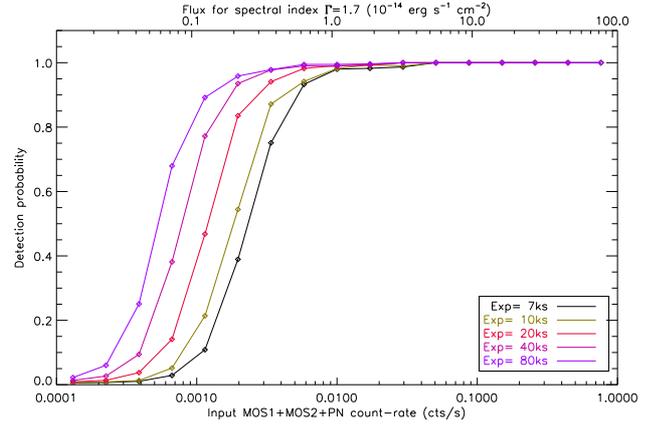}
 \caption{Efficiency of our source detection algorithm in the
   innermost annulus ($0-5$ arcmin) of the simulated soft band XMM
   observations, as a function of the input source count-rate (or
   equivalently, flux for a typical AGN spectrum and a galactic
   hydrogen column density fixed to $2.6\times10^{20}$ cm$^{-2}$). The
   exposure time differs from one curve to the other, but not the
   background rate.}
 \label{fig_probatexp} 
\end{figure}

\begin{figure}
	\includegraphics[width=91mm]{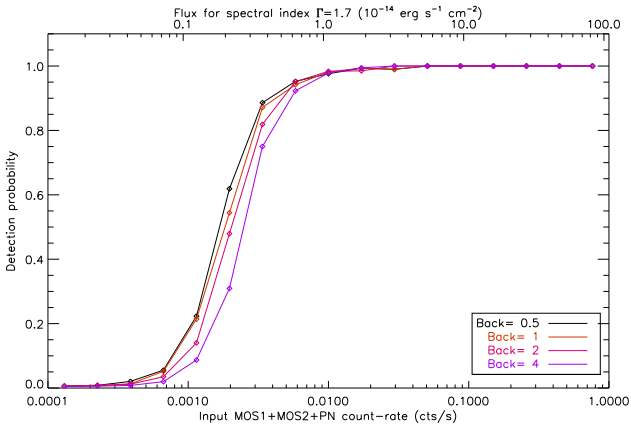}
 \caption{Same as Fig. \ref{fig_probatexp} for various background
   rates in the soft band (defined by a multiplicative factor times the values quoted
   in the 2$^{nd}$ column of Table \ref{table_background}). Exposure time is held at 10 ks in
   all cases.}
 \label{fig_probaback} 
\end{figure}

\begin{figure}
	\includegraphics[width=91mm]{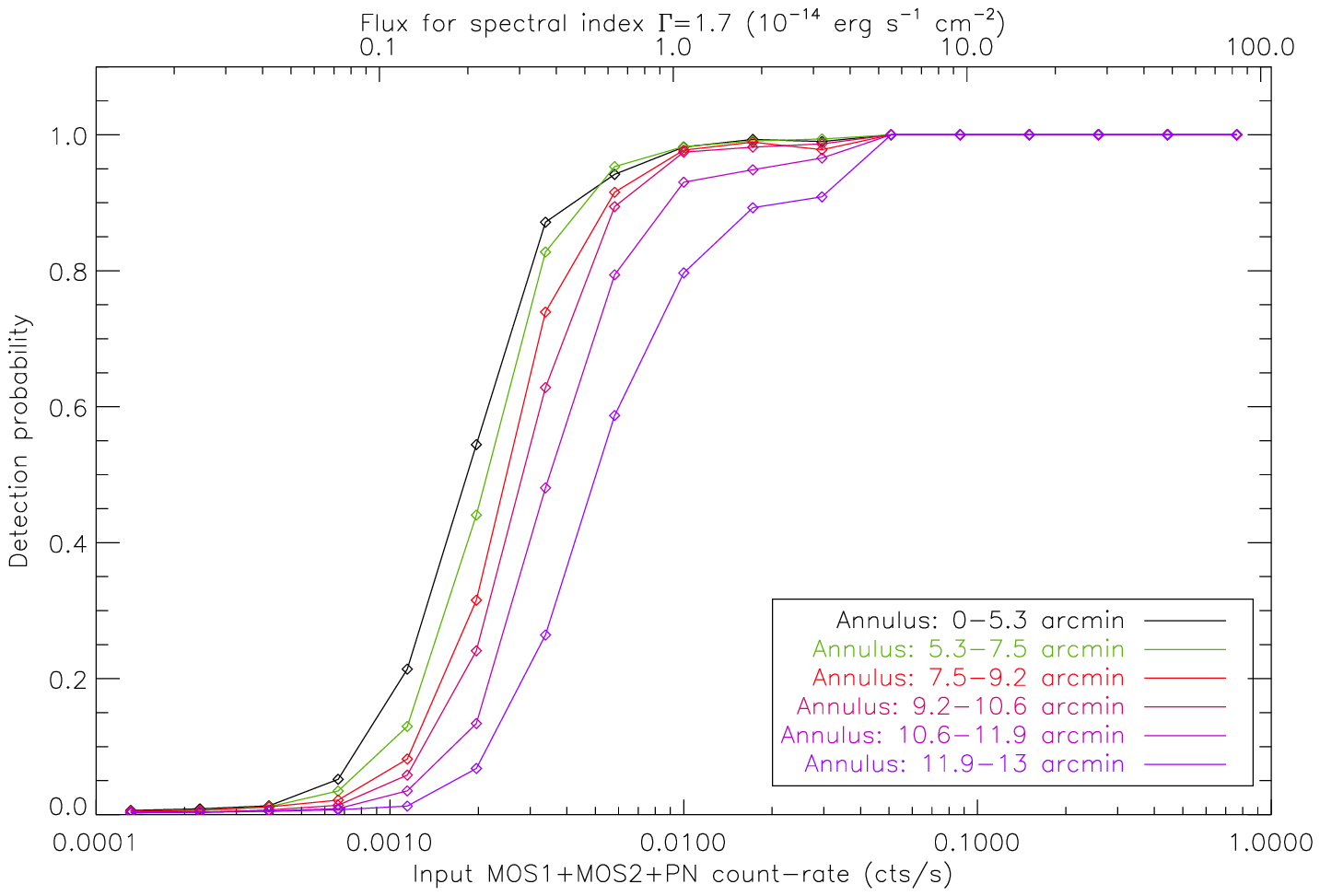}
 \caption{Same as Fig. \ref{fig_probatexp} for the six off-axis bins in the soft band
   (see text). Exposure time is held at 10 ks in all cases, and the
   particle background ratio is set to 1.}
 \label{fig_probaoffaxis} 
\end{figure}

We show in Figs. \ref{fig_probatexp}, \ref{fig_probaback} and
\ref{fig_probaoffaxis} the probability curves derived from our
simulations.
These curves were computed by dividing the number of detected (ML$>15$)
sources by the number of input sources in a given input count-rate bin
and for a given exposure time, background ratio and off-axis bin. The
detection efficiency is close to the flux-limited efficiency, whose
limit depends on the local pointing characteristics. A strong
dependence on the off-axis position is noticeable in
Fig. \ref{fig_probaoffaxis} because the effect of vignetting and PSF
distortions are growing with off-axis distance. The exposure time
dependence (Fig. \ref{fig_probatexp}) is compatible with a $\propto
\sqrt{T_{\mathrm{exp}}}$ improvement factor over the signal-to-noise ratio,
while the background level has a milder influence on the detection
efficiency.
In a typical pointing ($T_{\mathrm{exp}}=10~$ks, background ratio $=1$)
the flux limit is $2.5\times10^{-15}$ ($4\times10^{-15}$) erg s$^{-1}$
cm$^{-2}$ at 50\% (90\%) completeness.

\subsubsection{Relating real data to simulations}
\begin{figure*}
	\includegraphics[height=70mm]{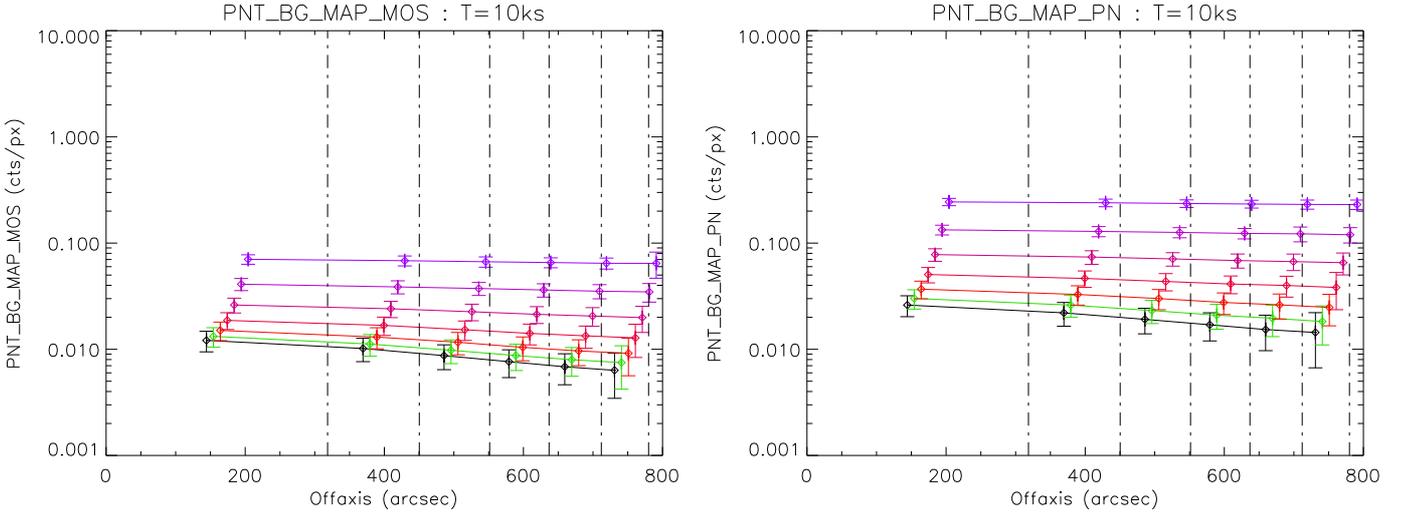}
 \caption{Pipeline-estimated background values on MOS (left) and PN
   (right) detectors, in the soft band, as seen in the simulations.
 Curves from bottom to top stand for background ratios $B$ equal to 0.1,
 0.25, 0.5, 1, 2, 4 and 8.
 This plot shows how the local background estimate output of the
 detection algorithm can be related to the background ratio parameter
 introduced in the simulations.
 The error bars represent 1-$\sigma$ standard variation computed from the source sample.
 The vertical lines correspond to bounds of our equal-area off-axis bins. 
 Only the results for 10 ks are displayed and similar relations are
 extracted for 7, 20, 40 and 80 ks pointings.}
 \label{fig_backrelation} 
\end{figure*}

Sensitivity maps across the entire XMM-LSS field can be derived
through interpolation between simulated pointings.
The exposure time of a given pointing is a straightforward quantity,
as is the off-axis angle at the position of a source.
To relate the background ratio quantity to real data, we used
estimates of the local background fitted by our detection algorithm at
each detected source position (see Pacaud et al. 2006 for a
description of the fitting procedure). Estimated numbers of background
counts per pixel are put out as two quantities PNT\_BG\_MAP\_MOS and
PNT\_BG\_MAP\_PN. Fig. \ref{fig_backrelation} shows the relationship
between the input background ratio and these quantities as derived from
simulations.
As expected, local background estimates computed by the detection
algorithm are well correlated with the background ratio values introduced
in the simulations. We fitted the local background values by PNT\_BG\_MAP\_MOS(PN) 
using the least-squares method separately for each pointing.
With this we determined the background ratio level $B$, which
corresponds to the best parameter of the fitting.

\section{Sky coverages and  log $N-$log $S$ distributions}

\begin{figure} 
\includegraphics[width=96mm]{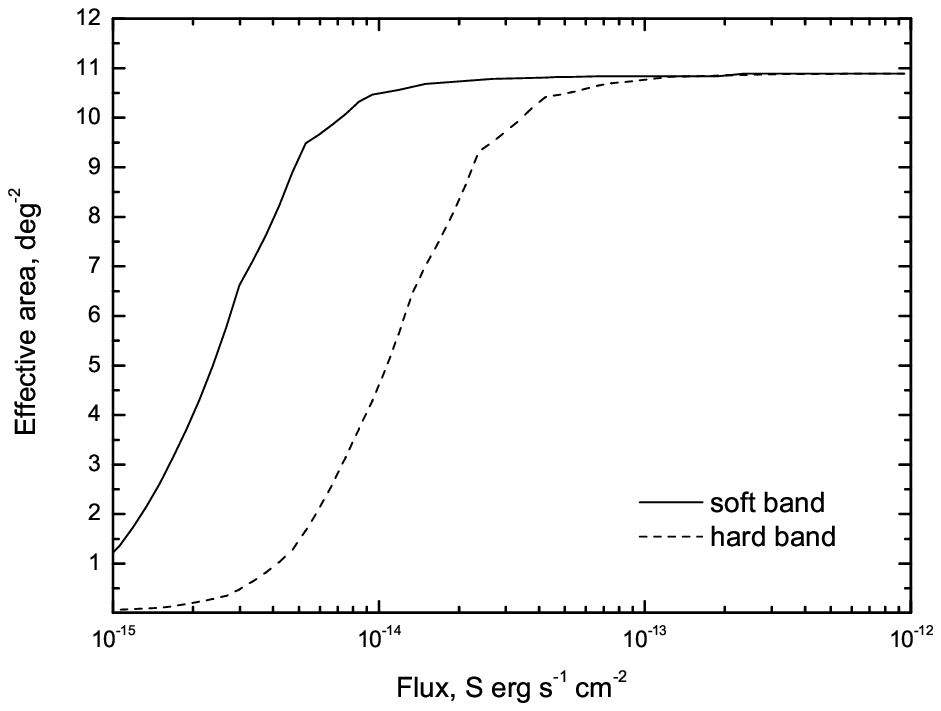}
\caption{Effective area curves for the whole XMM-LSS field %plus
%  Subaru) and for the Subaru subsample 
  in the soft (0.5-2 keV) and hard (2-10 keV) bands.}
\label{fig:effar}
\end{figure}

An important characteristic of an X-ray survey is the sky coverage or,
in other words, the effective area curve. This indicates the
maximum effective area over which we can detect sources brighter than
some given flux limit. We have constructed the area curves using the
numerically calculated probabilities $p$ to detect sources with a
certain flux $S$, an off-axis distance $R$ in a pointing with some
effective exposure $T$ and particle background level $B$. The
effective area $A(S)$ is calculated while integrating over the whole
field area $\Omega $:
\begin{equation}
A(S)=\int p(S,R,T,B) d\Omega. 
\end{equation} 
Fig.~\ref{fig:effar} shows the effective area curves for the
investigated samples in the soft and hard bands, with a minimum flux
$10^{-15}$ for the soft and $3 \times 10^{-15}$ erg s$^{-1}$ cm$^{-2}$ 
for the hard bands. For the construction
of the illustrated area curves we used the Voronoi tessellation
delimitation method.

\begin{figure} 
\includegraphics[width=96mm]{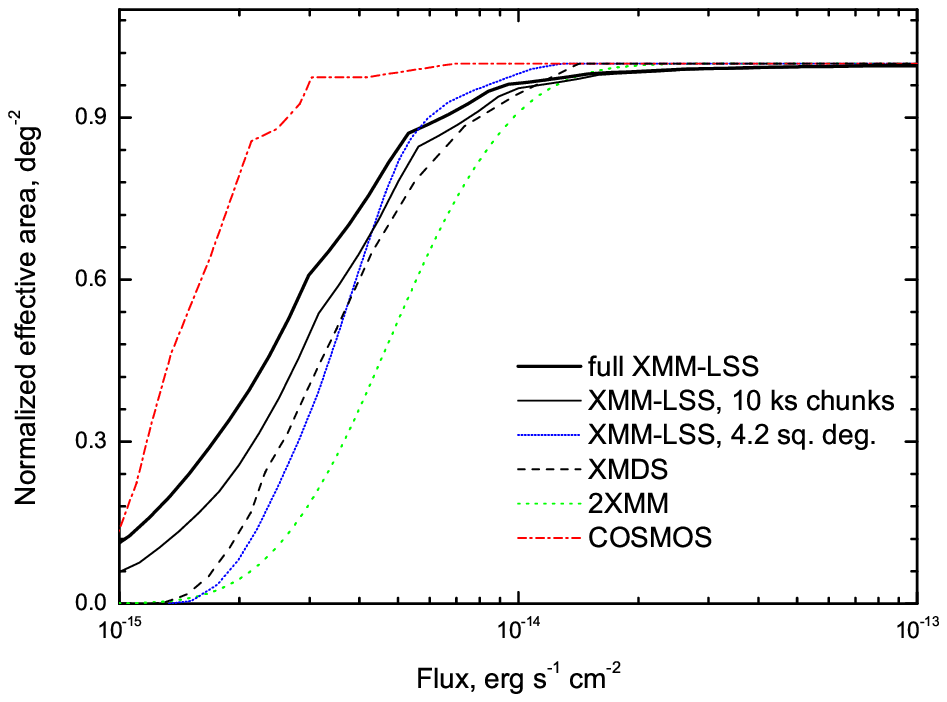}
\caption{Comparison of the normalized effective area curves
    in the soft band for the full exposure XMM-LSS field, the 10 ks version (see
    subsection 5.1), the XMM-LSS 4.2 sq. deg. \citep{gandhi06}, the XMM
    Medium Deep Survey \citep[XMDS][]{chiappetti05}, the 2XMM
    \citep{ebrero} and the COSMOS \citep{Miyaji_2007, Cappelluti07}.}
\label{fig:ac}
\end{figure}

 Fig.~\ref{fig:ac} shows a comparison between the normalized
  effective area curves of various recent X-ray surveys. The effective
  area curve as a function of flux depends mainly on the depth of the
  source detection (indicated by the signal-to-noise ratio 
  or likelihood thresholds). It
  also depends on the distributions of the pointing exposures,
  particle background level and the procedure of handling the
  pointing overlaps. Evidently, the COSMOS field has the lowest
  flux limit and the steepest area curve among the
  considerable surveys with the likelihood limit for the source detection
  being equal to 6 \citep{Cappelluti07}. Our full exposure XMM-LSS
  survey, having a significant fraction of the contributing pointings
  with exposures between 10 and 15 ks and a source detection
  threshold of ML$=15$, has the next lowest flux-limit after the
  COSMOS survey, and a quite steeply increasing area-curve.
  The corresponding 10 ks XMM-LSS field has its area curve shifted
  to the right and its flux limit increased by a factor of $\sim 1.2$.

Using those area curves and the differential distributions of the
sources as a function of their flux, we constructed the
 log $N-$log $S$ relation. Note that it is important to take into account
the flux boosting. This phenomenon especially affects faint objects
with a low detection probability. Owing to Poisson noise, we may
detect objects fainter than the flux limit in successful cases
and sometimes not detect sources brighter than the flux limit in
unsuccessful cases. This may cause the creation of an
artificial bump in the  log $N-$log $S$ distribution.

\begin{figure} 
\includegraphics[width=96mm]{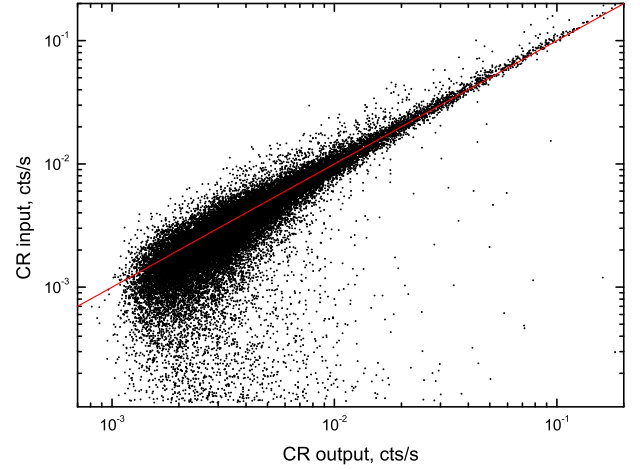}
\caption{Example of the dependence between input and output
  count-rates for the hard band, $T=20$ ks, $b=1$. The red line corresponds to $CR_{in}$=$CR_{out}$. }
\label{fig:inout}
\end{figure}

\begin{figure} 
\includegraphics[width=96mm]{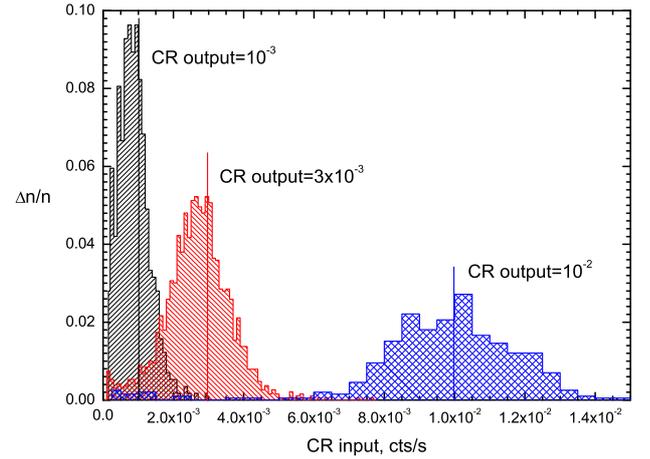}
\caption{Distributions of $CR_{in}$ for three detected $CR_{out}$
  created on the basis of the simulated distribution in Fig. 11.}
\label{fig:dist3}
\end{figure}

\begin{figure} 
\includegraphics[width=\columnwidth]{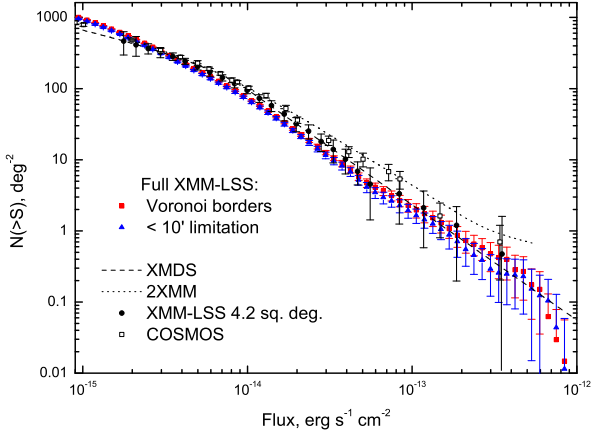}
\caption{Log $N-$log $S$ distributions in the soft band for the whole
  XMM-LSS sample and for the two different procedures of handling the
  pointing overlaps. The results
    of the XMM Medium Deep Survey (XMDS) \citep{chiappetti05}, 2XMM
    \citep{ebrero}, XMM-LSS 4.2 sq. deg. \citep{gandhi06} and COSMOS
    \citep{Miyaji_2007, Cappelluti07} are shown for
    comparison. The vertical bars denote $1\sigma $ uncertainties.}
\label{fig:lnls_s}
\end{figure}

\begin{figure} 
\includegraphics[width=\columnwidth]{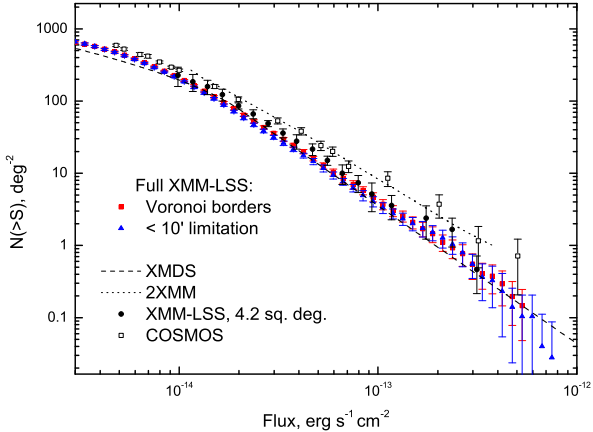}
\caption{Log $N-$log $S$ distributions in the hard band for the whole
  sample and for the two different procedures of handling the
  pointing overlaps. For
    comparison we present the $\log N-\log S$ distributions for the same samples
    as in Fig. \ref{fig:lnls_s}.}
\label{fig:lnls_h}
\end{figure}

To take this effect into account, we used the numerically
simulated dependencies between the input $CR_{in}$ and the output
$CR_{out}$ count rates individually for each pointing (see
Fig.~\ref{fig:inout} for example). Clearly, when we detect some
flux $CR_{out}$, it corresponds to a real input $CR_{in}$ distributed
over a wide range.
For each $CR_{out}$ bin we constructed the density probability
distribution as a function of $CR_{in}$. 
Fig.~\ref{fig:dist3} represents the normalized distributions of
$CR_{in}$ for three detected $CR_{out}$.
At low flux, we may see an asymmetric shape in the distribution that
is shifted toward smaller $CR_{in}$ because of an artificial flux
boosting.
Therefore, we randomly chose some $CR_{in}$ for each detected source with $CR_{out}$ according to the density probability function.
In this way, we carried out  Monte-Carlo simulations with the
deconvolution of the output into the input rates and constructed
 log $N-$log $S$ curves for various considered samples
(Figs.~\ref{fig:lnls_s}~-~\ref{fig:lnls_h}). 

The currently estimated $\log N-\log S$ are
lower for both bands than those of the 2XMM \citep{ebrero} and COSMOS \citep{Cappelluti07} surveys, with
  deviations not exceeding the $2-3\sigma$ Poisson level. However, they excellently agree with those derived by  \cite{gandhi06} and \cite{chiappetti05}, based on previous releases of
  XMM-LSS fields. Moreover, the XMDS \citep{chiappetti05} was based on a totally different pipeline used for extracting the X-ray point-like sources.
 This suggests that the observed
  deficiency could be an intrinsic characteristic of the XMM-LSS field.

\section{The angular correlation function analysis} 

To determine the ACF, we generated random catalogs in the following way. Firstly, we
distributed the fiducial point-like sources with random coordinates
over the whole investigated field. Secondly, we
chose for each random source a flux according to the log $N-$log $S$ distribution and
calculated the probability $p$ of detecting the corresponding
point-like source in the relevant pointing, taking into account the
exposure time, the particle background level ($B$) of the pointing and
the off-axis distance of the corresponding source. Then, we chose a random number $\rho $ for each
random point-like source that is uniformly
distributed between 0 and 1. If the $\rho $ value was less than $p$, we
kept the source, if it was higher, we discarded the source. 
If a random source was closer than 10 arcsec to another one, we removed it
because the extension of the EPIC PSF ($\sim 6^{''}$ minimum, on
  axis) prevents one from detecting such close pairs and blends them into a
  single source.
We generated random catalogs in this way that contain 100 times the
number of point-like sources in the real source catalog, that was 
used in the present analysis. The larger the point population of
  the random catalog, the more accurate the ACF measurement because it suppresses
  random fluctuations caused by small numbers.

To calculate the ACF, 
%we have followed the approach described in \cite{gandhi06}. Namely, 
we used two estimators, the Hamilton estimator
\citep{hamilton}, as in \cite{gandhi06}:
\begin{equation}
1+w(\theta )=f_H \frac{DD(\theta) RR(\theta)}{DR^2(\theta)}\;,
\end{equation}   
and the Landy \& Szalay estimator \citep{landy}:
\begin{equation}
1+w(\theta )=f_{LS} \frac{DD(\theta)-2 DR(\theta) + RR(\theta)}{RR(\theta)}\;,
\end{equation}   
where $DD$, $RR$ and $DR$ represent the numbers of data-data,
random-random and data-random pairs with a separation $\theta$,
 while $f_H$ and $f_{LS}$ are the corresponding 
normalization factors of the two estimators. In general the two
estimators provide consistent results but in any case we will present the results based on both estimators in
the correlation function plots.

To speed-up our calculations, we divided the random
catalog of those samples with more than 2000 X-ray sources
into a maximum of 10 random subcatalogs, and we averaged
$w(\theta)$ for each $\theta$ bin over the whole random
catalogs. Note that we verified by investigating one such sample that the
above procedure provides stable correlation results.
The $w(\theta)$ uncertainty in each $\theta$-bin is given by
\begin{equation}
\sigma_{w}=(1+w)/\sqrt{DD} \;.
\end{equation}

The ACF calculations were performed for angular scales in the range:
$20''<\theta<\theta_{\rm max}$, where $\theta_{\rm max}=12000''$.
% for the case of the whole XMMLSS+SUBARU sample.
% and $\theta_{\rm  max}=3000^{"}$ for the case of only the SUBARU
% field.
 We used $20''$ as our lowest angular-separation limit 
  because of the large size of the XMM-Newton PSF near the FoV borders.
 We verified that pairs, constituted by sources belonging to 
adjacent pointings, are real only for pair separations $\magcir 20''$.
 We then fitted the resulting ACF with the power-law in the
angular range where it was possible, i.e., using only the positive $w$ values:
\begin{equation}
w(\theta )=(\theta_{0} /\theta)^{\gamma -1}\;.
\end{equation}

\begin{figure}
\includegraphics[width=0.9\columnwidth]{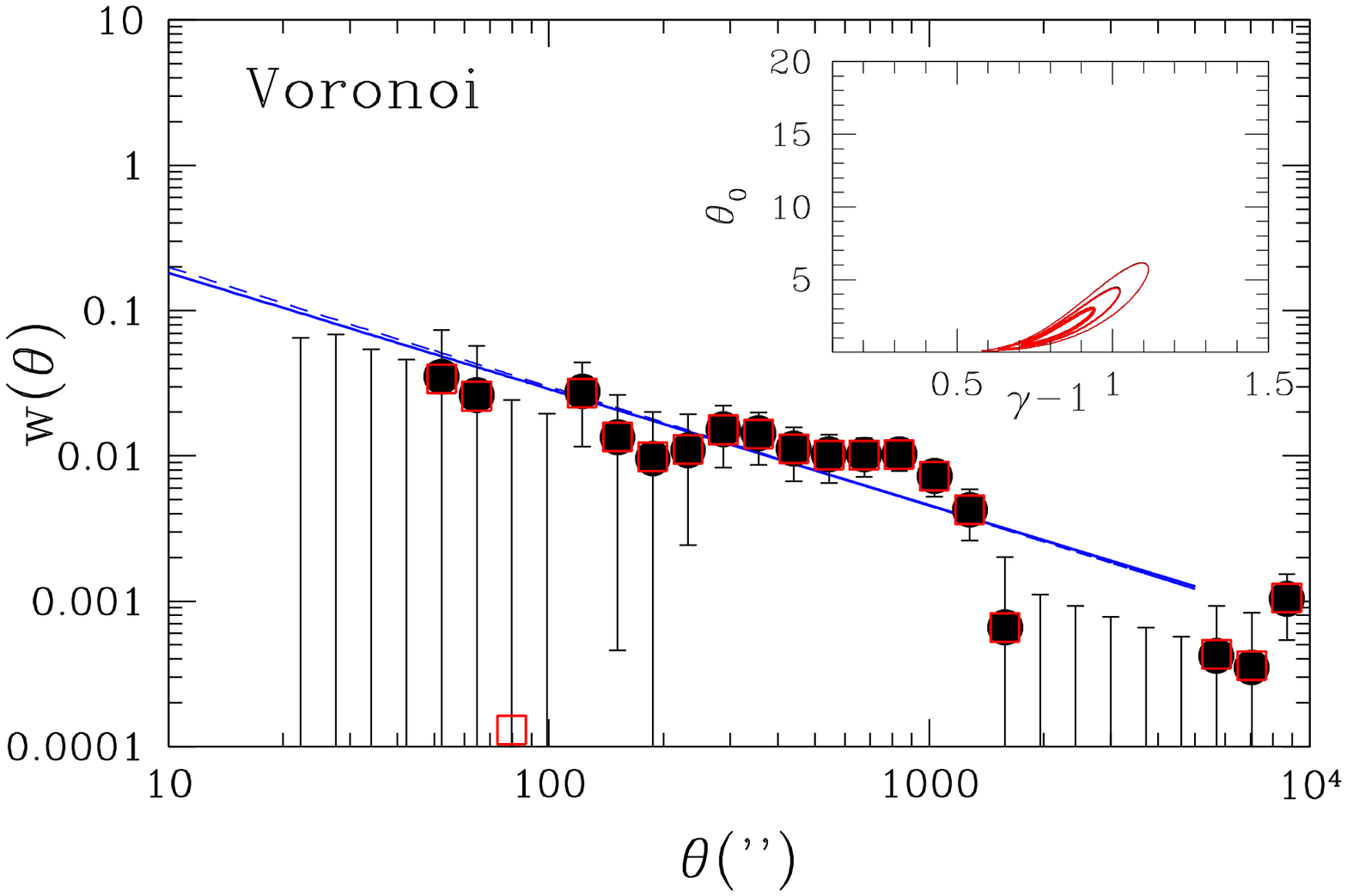} \hfill
\includegraphics[width=0.9\columnwidth]{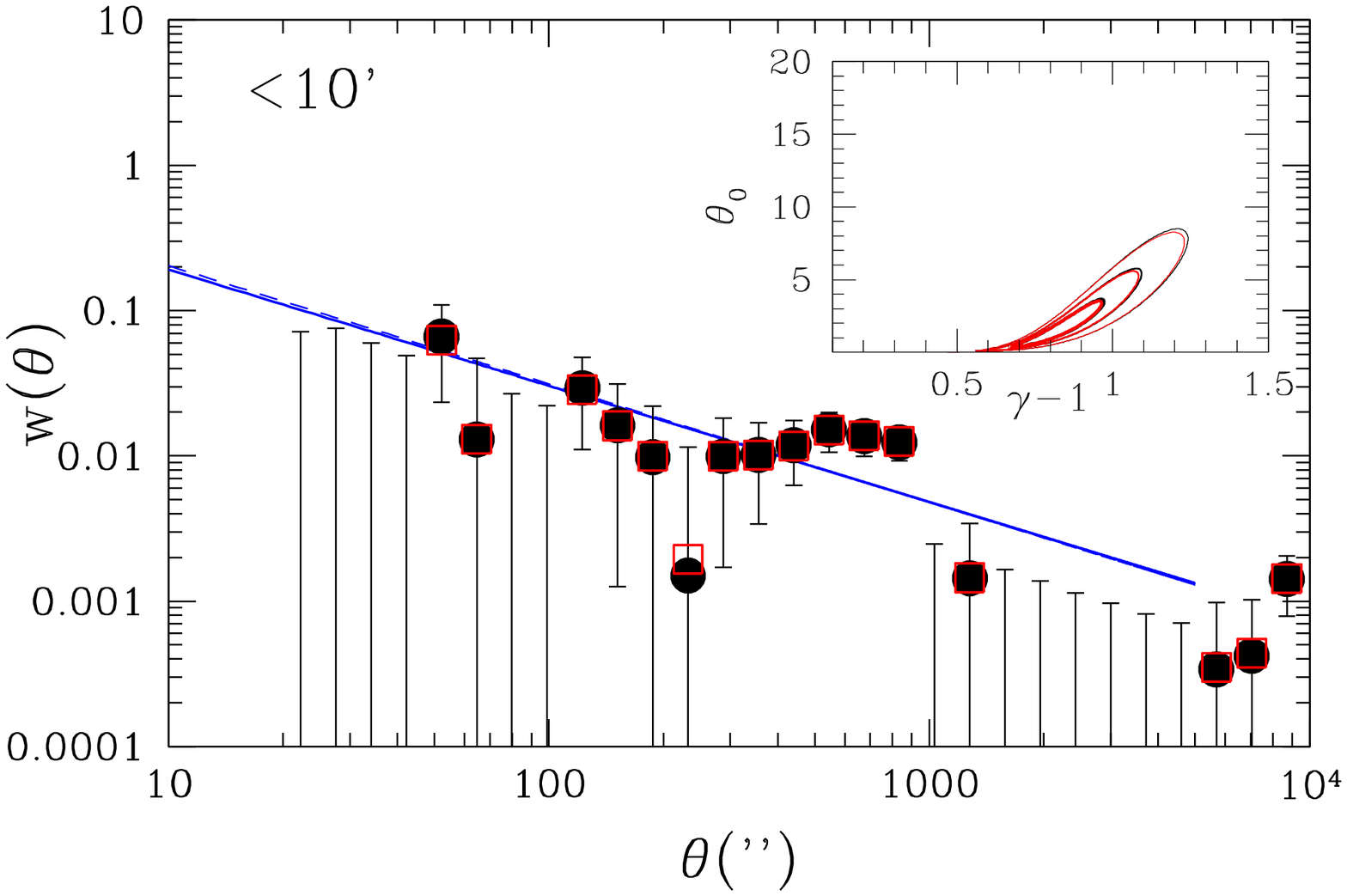}
\caption{Soft band ACF for the whole sample: Voronoi delimitation (top
  panel); off-axis angle $<10^{'}$ (lower panel). 
The filled (black) points correspond to the Hamilton estimator
while the open (red) squares to the Landy \& Szalay estimator. 
The error bars represent 1$\sigma$ standard deviation.
The dashed line represents the best power-law
  fit, while the continuous line corresponds to the constant $\gamma=1.8$
  fit. The inset plot presents the 1, 2 and 3$\sigma$ contours in the
  fitted ($\theta_0, \gamma$) parameter space.}
\label{fig:xsv}
\end{figure}
\begin{figure}
\includegraphics[width=0.9\columnwidth]{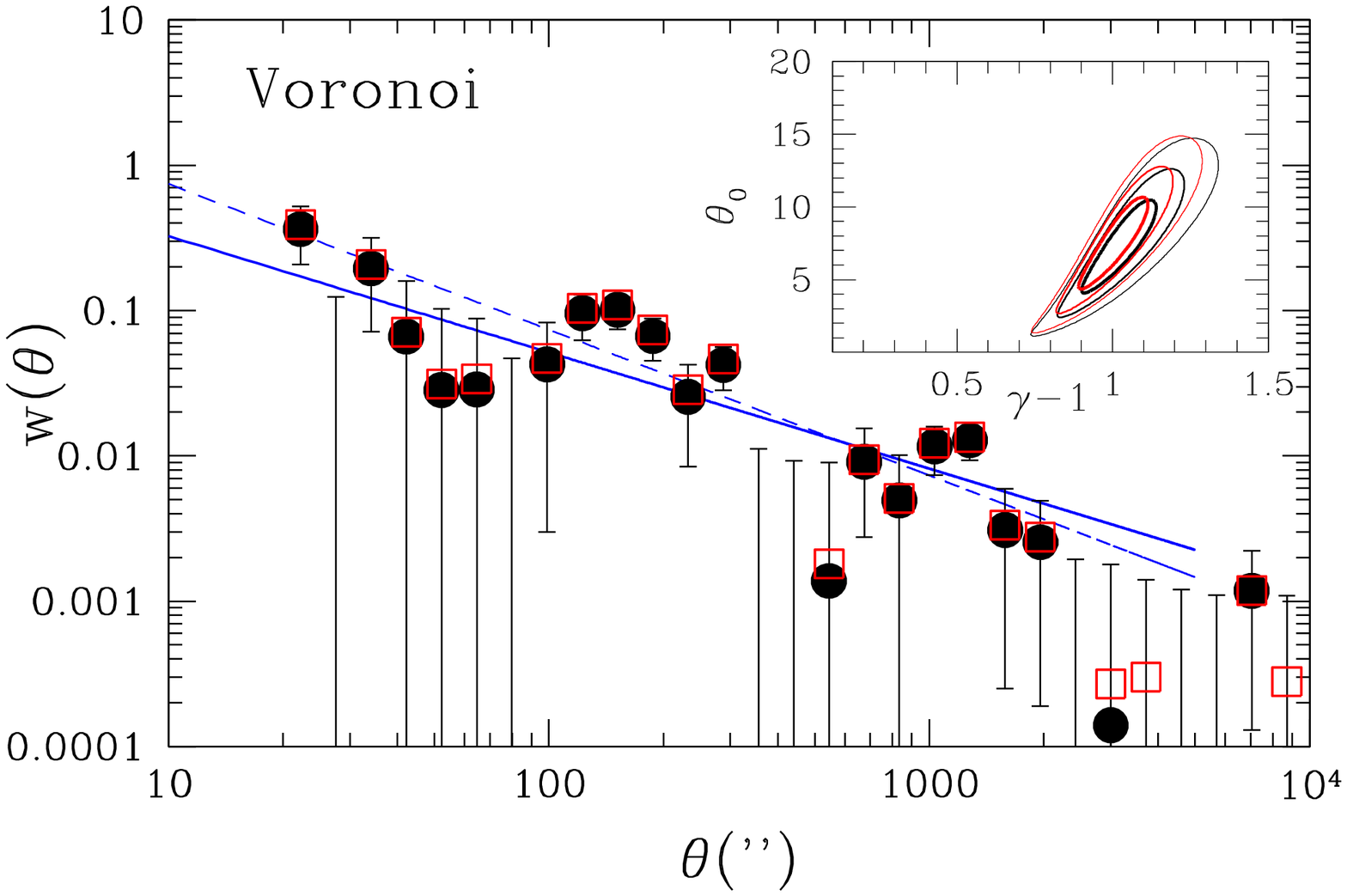} \hfill
\includegraphics[width=0.9\columnwidth]{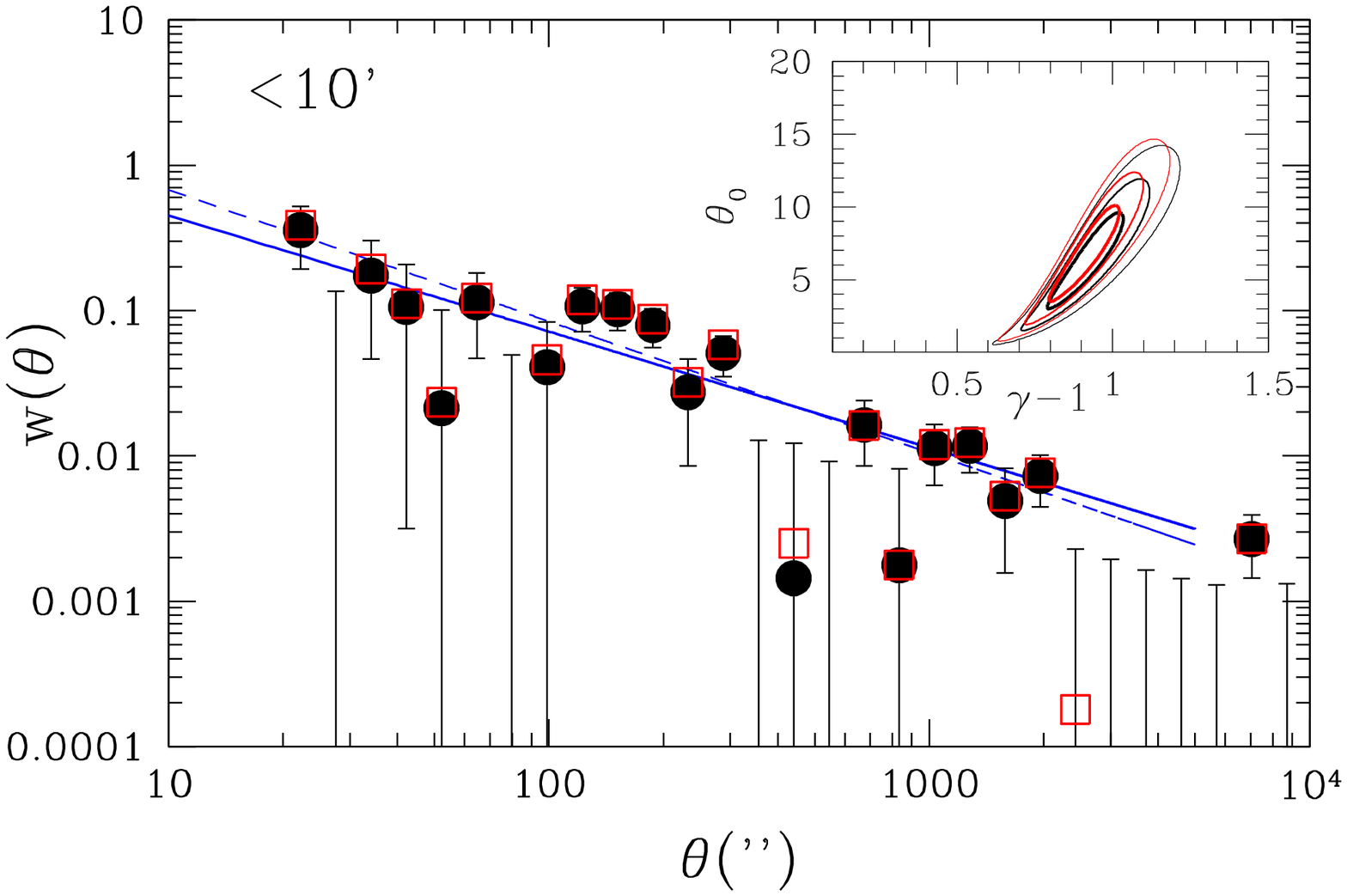}
\caption{Hard band ACF for the whole sample: Voronoi delimitation (upper
  panel); off-axis angle $<10^{'}$ (lower panel).}
\label{fig:xhv}
\end{figure}

We analyze here %separately 
the ACF of the full exposure XMM-LSS field.
 %which includes also the deeper field, ie., using all the available
 %exposure time. 
However, because there are strong indications for a flux-limit dependence of the
correlation function amplitude \citep[e.g.,][]{Plionis08}, we also
analyzed a homogeneous sample of an effective 10 ks exposure over 
the whole XMM-LSS region. To this end we cut the event list 
of the pointings into 10 ks chunks and
  repeated the source detection procedure from the beginning.
We also separately estimated the ACF %of the deeper SUBARU field, as well as  
of samples based on the hardness ratio (HR).

\subsection{The whole XMM-LSS field}
We first present in Figs.~\ref{fig:xsv} and ~\ref{fig:xhv}
the ACF results of the full exposure XMM-LSS region
for both the soft and hard bands and for the 
Voronoi delimitation and off-axis angle $<10'$ overlap approaches. In
the inset panels we present the 1, 2 and 3$\sigma$ contours of the
fitted parameters in the ($\theta_0, \gamma$) plane, 
 while in Table~\ref{tab:main} we present the corresponding best fit
$\theta_{0}$ and $\gamma$ parameters and their standard
deviation, as well as the value of $\theta_0$ for a fixed
slope $\gamma=1.8$ and the integral ACF signal within separations of 3.3
arcmin, $w(<3.3^{'})$. Evidently, that there are no
significant differences between the results based on the
correlation function estimators (as seen in Figs.~\ref{fig:xsv} -~\ref{fig:xhv}) or on the two
delimitation methods. Therefore we used for the
remaining study only the samples based on the space-filling Voronoi
delimitation method and the Landy \& Szalay ACF estimator (see also
\cite{Ker00} for a detailed comparison of different estimators).

Furthermore, we find that the hard band correlation function is slightly
but clearly stronger than the corresponding soft band, as can be also verified by comparing the
corresponding inset contour plots, which agree with the
results of \cite{basilakos_2005}, \cite{Puccetti} but disagree with those of \cite{ebrero}.

\begin{table}[tbh]
\begin{center}
\caption{The soft and hard band correlation functions for the 
whole XMM-LSS field, as well as for the two possible
  overlap approaches. $N$ indicates the number of X-ray sources
  in the corresponding sample, while the last column shows the
  integrated ACF signal, and its uncertainty, within $20''<\theta<200''$.}
\tabcolsep 2pt

\begin{tabular}{ccccccc}
Band & Overlap & $N$ & $\theta^{''}_{0}$ & $\gamma$
&$\theta^{''}_{0,\gamma=1.8}$ & $w(<3.3^{'})$ \\
\hline
%soft & Vor. & 5094 & $0.64\pm 0.13$ & $1.73\pm 0.02$ \\
%     & $<10'$  & 4067 & $0.23\pm 0.08$ & $1.63\pm 0.03$ \\
Soft & Vor.   & 5093 & $1.3\pm 0.2$ & $1.94\pm 0.02$ & 1.3$\pm0.2$ & 0.006$\pm0.007$\\
     & $<10'$ & 4066 & $1.4\pm 0.3$ & $1.81\pm 0.02$ & 1.3$\pm0.2$ & 0.009$\pm0.003$ \\
\hline
%hard & whole & Voronoi & 2370 & $6.22\pm 1.09$ & $1.96\pm 0.04$ \\
% & whole & $<$10' & 1989 & $8.13\pm 1.31$ & $1.97\pm 0.04$ \\
Hard & Vor.   & 2369 & $7.5\pm 0.9$ & $2.00\pm 0.03$ & 2.5$\pm 0.4$ & 0.075$\pm0.013$\\
     & $<10'$ & 1988 & $6.5\pm 0.8$ & $1.91\pm 0.03$ & 3.7$\pm 0.5$ & 0.080$\pm0.014$\\ \hline
\end{tabular}
\label{tab:main}
\end{center}
\end{table}

\begin{figure}
\includegraphics[width=0.9\columnwidth]{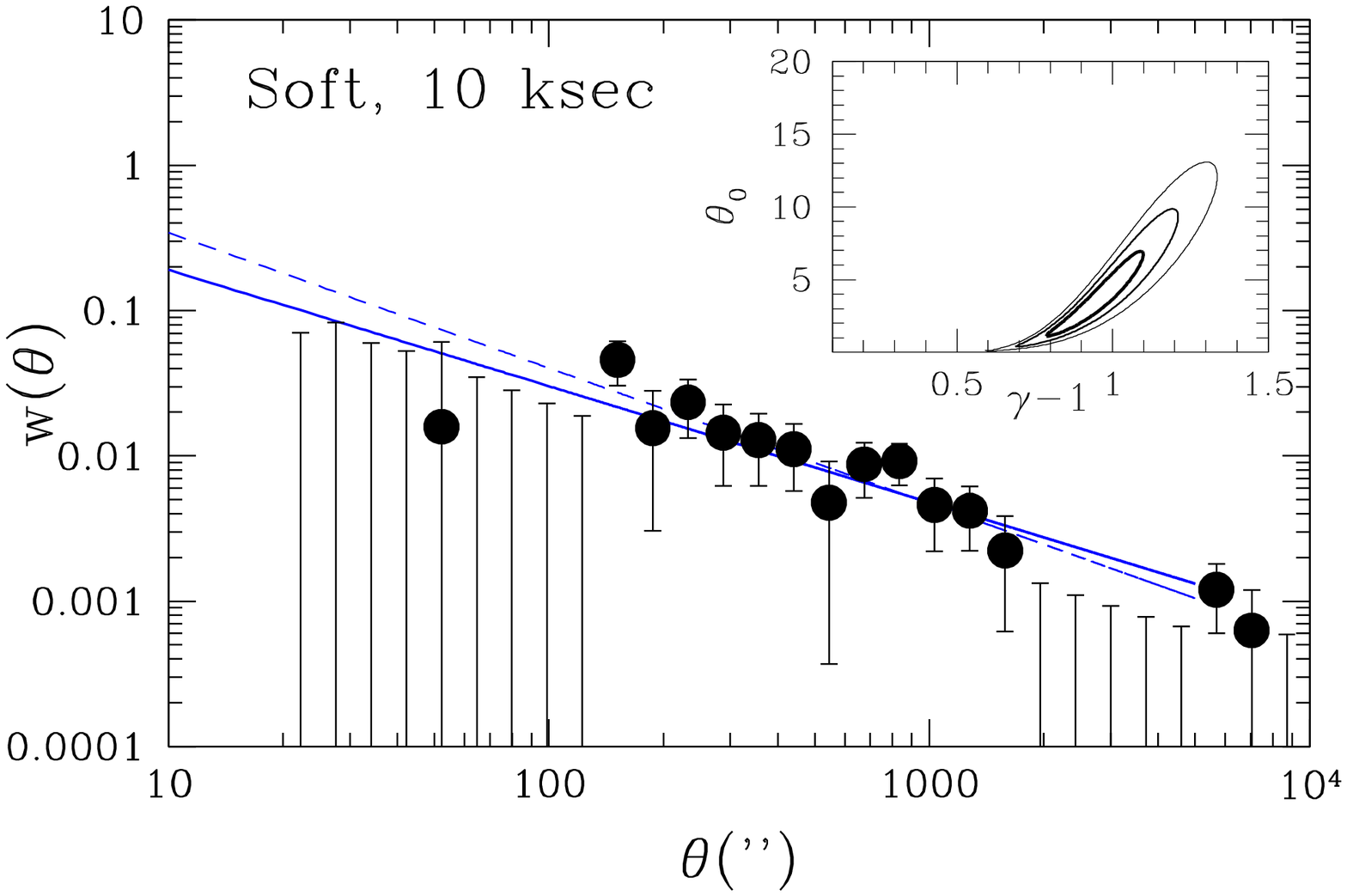} \hfill
\includegraphics[width=0.9\columnwidth]{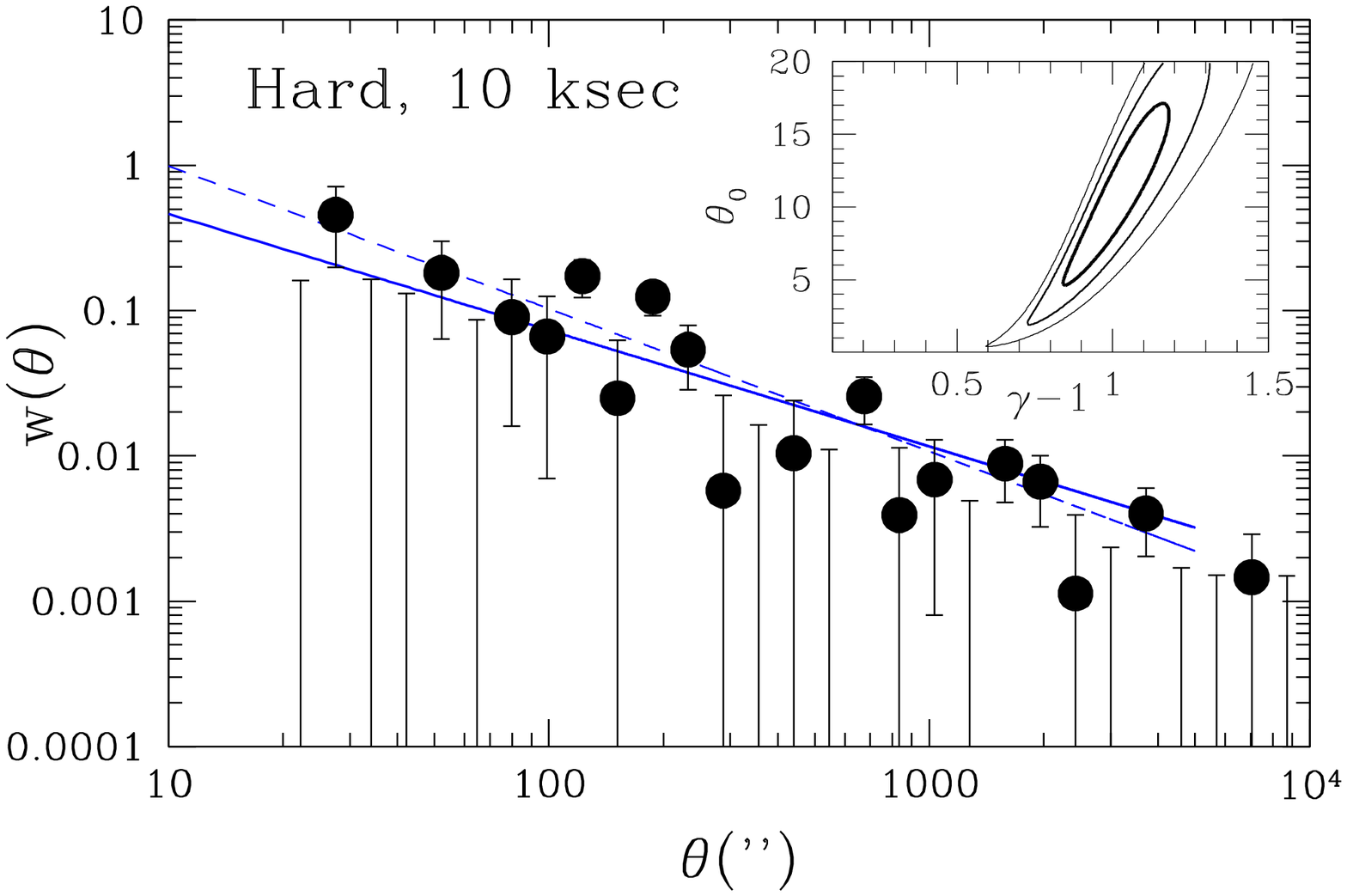}
\caption{ACF for the 10 ks sample in the soft band (upper panel) and
  for the hard band (lower panel).}
\label{fig:10ks_s}
\end{figure}

As discussed above, to provide a ``clean'' ACF, that is
unaffected by the convolution of (a) 
the variable flux-limit in the different parts of the survey 
and (b) the flux-limit clustering dependence,
we considered a sample with a homogeneous 10 ks
 exposure time across the whole XMM-LSS area. Table~\ref{tab:10ks}
and Fig.~\ref{fig:10ks_s} show the parameters of
the ACFs for both bands. The main variation with respect to the
previous analysis is that the ACF difference between the soft and hard
bands is now even more prominent.

\begin{table}[tbh]
\begin{center}
\caption{Correlation function for the 10 ks chunk samples.}
\label{tab1}
\tabcolsep 3pt
\begin{tabular}{ccccccc}
Band & $N$ & $\theta^{''}_{0}$ & $\gamma $ & $\theta^{''}_{0, \gamma=1.8}$ & $w(<3.3^{'})$\\
\hline
%soft & 4360 & $3.82\pm 0.46$ & $1.97\pm 0.02$ \\
%hard & 1712 & $14.07\pm 1.91$ & $2.18\pm 0.05$ 
Soft & 4360 & $3.2\pm 0.5$ & $1.93\pm 0.03$ & 1.3$\pm 0.2$ & 0.005$\pm0.007$\\
Hard & 1712 & $9.9\pm 1.4$ & $1.98\pm 0.04$ & 3.8$\pm 0.7$ & 0.092$\pm
0.019$\\ \hline
\label{tab:10ks}
\end{tabular}
\end{center}
\end{table}

\begin{figure}
\includegraphics[width=0.9\columnwidth]{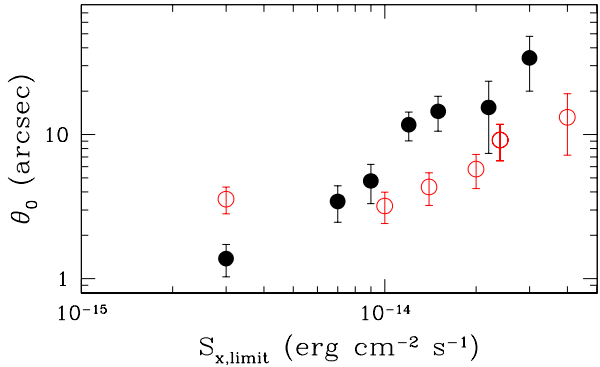}
\caption{Best-fit correlation length $\theta _{0}$ for $\gamma =1.8$
  as a function of the flux limit of the homogeneous 10 ks sample in the soft (filled
  circles) and the hard bands (open circles).}
\label{fig:limit}
\end{figure}

We also investigated the flux-limit dependence of clustering with our homogeneous 10 ks sample. To this end we estimated the
angular clustering length, $\theta_0$ for various flux-limited
subsamples by keeping the slope of the ACF fixed to its nominal value
of $\gamma =1.8$. Fig.~\ref{fig:limit} shows the corresponding 
results for the soft and hard bands. 
Evidently the known dependence is clearly reproduced with our
data, and it will be interesting to investigate whether this
dependence is present in the spatial correlation length, via
Limber's inversion (see further below). Another interesting result is
that the amplitude of the hard band ACF is larger than that of the
soft band only in the lowest flux-limits. At flux limits $\geq
10^{-14}$ erg s$^{-1}$ cm$^{-2}$ the trend is reversed and the soft
band is stronger than the hard band clustering.

How do our results compare with those of other XMM surveys ? With
respect to our previous release of the 4.2 sq. deg. XMM-LSS survey
\citep{gandhi06}, our new catalog introduces many
improvements. Among them is the wider (by $\sim 2.6$ times) sampled
area, and the inclusion of a deeper SXDS field.
%, which finally result in a significantly larger (by $\sim
%5$ times) X-ray point-like source population.
Furthermore, we updated the point-like source detection
procedure and introduced a novel definition of the 
selection function and random-catalog generation procedure.

\begin{figure}
\includegraphics[width=1.1\columnwidth]{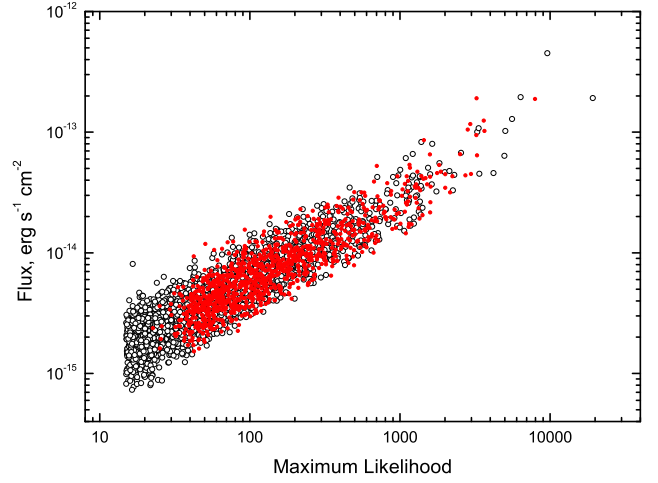}
\caption{Dependence between flux and ML in the soft band
    for the point like sources of 44 pointings from the present XMM-LSS
    survey (black open circles) and from \cite{gandhi06} (red filled
    circles).}
\label{fig:flux_ml}
\end{figure}

Our current XMM-LSS area curve is substantially different from that of
\cite{gandhi06} (see Fig.~\ref{fig:ac}). To investigate the reasons of this difference in detail, we used the 44 pointings common to both studies to compare the corresponding point-like source catalogs in the soft band.
The current XMM-LSS catalog contains 2106 objects with off-axis distances less than
$10'$ and the \cite{gandhi06} catalogue contains 1093 such sources, while the
common sources are 1048.
Fig.~\ref{fig:flux_ml} shows the dependence between flux and ML for
both catalogs. Obviously, the chosen ML limit of the \cite{gandhi06} catalog is
substantially higher than the current limit of ML $=15$ and it is equal to ML $\sim 40$. 
%Fig.~\ref{fig:sn_ml} represents the comparison of signal to noise
%ratio and ML by \cite{gandhi06}. We can conclude that used there limit
%$S/N~>~3$ corresponds roughly to ML$~>~40$.
%So the difference between these two catalogs is in the source
%detection threshold: ML$=15$ for present work and ML$\sim 40$ for the
%case of the \cite{gandhi06} catalog.
It is also evident, inspecting Fig.~\ref{fig:flux_ml}, that a 
value of ML $\sim 40$ is associated with a significantly higher
flux-limit with respect to that of ML $=15$, causing the observed difference
of the corresponding area curves 
% at fluxes below $10^{-14}$ erg s$^{-1}$ cm$^{-2}$ 
(see Fig.~\ref{fig:ac}).
%range between $\sim %10^{-15}$ and  $\sim 7\times 10^{-15}$ erg s$^{-1}$ cm$^{-2}$ (see

%\begin{figure}
%\includegraphics[width=1.1\columnwidth]{sn_ml.eps}
%\caption{{\bf The dependence between signal to noise ratio and ML in
%the soft band for the point like sources from 44 pointings from
%\cite{gandhi06}}}
%\label{fig:sn_ml}
%\end{figure}

The above mentioned changes and improvements, particularly the lower
ML detection limit, have resulted in a variation of some of
our results with respect to those of \citet{gandhi06}. 
Specifically, we found a slightly different
clustering signal in the soft band; $\theta_{0}=1.3''\pm 0.2''$ for $\gamma= 1.81$ vs
$6.3'' \pm 3''$ for $\gamma=2.2$ in \cite{gandhi06}. 
However, at the fixed canonical value of the exponent ($\gamma = 1.8$), the \cite{gandhi06} soft band analysis provides a clustering amplitude of $\theta_{0} = 1.7''\pm 0.9''$ versus $1.2''\pm 0.2''$ for the current XMM-LSS survey. The lower correlation signal of our current XMM-LSS survey
  should be attributed to the lower ML limit, which
introduces a significantly higher fraction of faint sources with
respect to the higher ML limit of the \cite{gandhi06} catalog.
Also, we found a significant clustering signal in the hard band, in contrast to the absence of any significance in \cite{gandhi06}.

With respect to the COSMOS \citep{Miyaji_2007} and 2XMM \citep{ebrero} surveys,  
we find (at fixed canonical $\gamma =1.8$) 
a lower soft band correlation function amplitude, 
$\theta_{0}=1.2''\pm 0.2''$, compared to $1.9''\pm 0.3''$ and to
$7.7''\pm 0.1''$ for the COSMOS and the 2XMM surveys, respectively.
Our hard band (2-10 keV) XMM-LSS correlation amplitude of $3.6''\pm 0.7''$ is also lower
than the corresponding 2XMM value of $5.9''\pm 0.3''$, while the COSMOS
hard band correlation results are not very significant, probably
because they are divided into two sub-bands (2-4.5 and 4.5-10 keV).

Note, however, that the wide contiguous area of the XMM-LSS survey implies
that we should have a better estimation of $w(\theta)$ on large
angular scales (ie., 1000$''\mincir \theta \mincir 10000''$), 
while COSMOS and 2XMM are limited to $\sim 6000''$ and $\sim 1000''$, respectively.

\begin{figure}
\includegraphics[width=0.9\columnwidth]{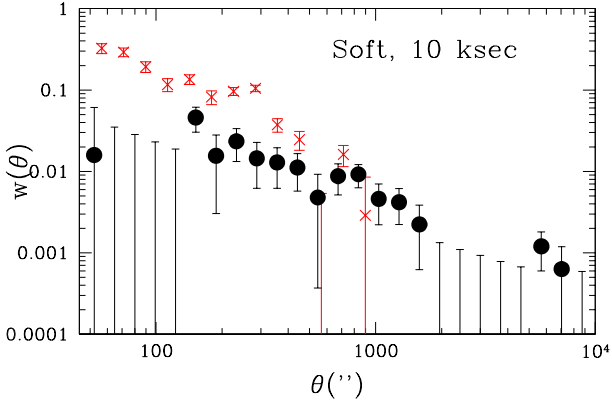}
\caption{Soft band ACF comparison between our XMM-LSS survey 
  (filled points) and that of the 2XMM survey (crosses) of \citet{ebrero}.}
\label{fig:comp}
\end{figure}

In Fig. \ref{fig:comp} we compare the soft band $w(\theta)$ of our XMM-LSS 
and the 2XMM surveys. The large 2XMM $w(\theta)$ amplitude at small
angular scales is evident, although at $\sim 1000''$ the two
correlation functions appear to be consistent.
The higher 2XMM correlation amplitude should be attributed
to the considerably different mix of faint and bright sources in the
two surveys, as shown by their respective area curves
(see Fig.~\ref{fig:ac}). The larger part of faint sources in the
current XMM-LSS survey causes the lower amplitude of the source angular
correlation function with respect to the 2XMM, as expected from the
known dependence between clustering and flux-limit \citep{Plionis08,
  ebrero}, a fact which has also been verified by our analysis
(Fig.~\ref{fig:limit}).

\subsection{Subsamples of sources with soft and hard spectra}
An interesting question, that relates to the unification paradigm of AGN,
is whether the clustering pattern, among others, of hard and
soft-spectrum AGN is comparable. According to the unification paradigm,
what determines the appearance of an AGN as obscured or unobscured
(type II or I) is its orientation with respect to the observer's line-of-sight. Therefore, there should be no intrinsic difference in their
clustering pattern. On this question there have been conflicting results
in the literature and we here re-address this with our data.

\begin{figure} 
\includegraphics[width=\columnwidth]{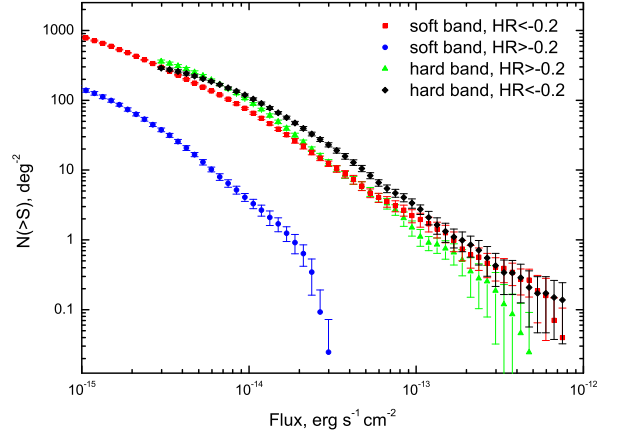}
\caption{Log $N-$log $S$ distributions in the soft and the hard bands
  for sources with different hardness ratios.}
\label{fig:lnls_hr}
\end{figure}

\begin{figure}
\includegraphics[width=0.9\columnwidth]{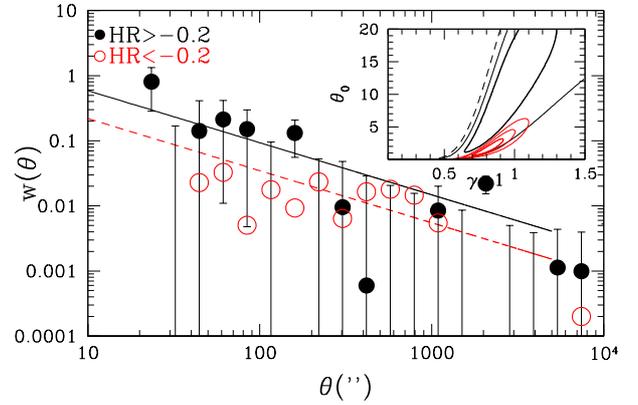}
\caption{ACF for the whole XMM-LSS sample in the soft
  band for sources with HR$>-0.2$ (filled circles, hard-spectrum AGN), and for
  sources with HR$<-0.2$ (open circles; soft-spectrum AGN). Note that
  for clarity reasons we do not plot the $w(\theta)$ uncertainties of
  the later sources. The solid
  line represents the $\gamma=1.8$ fit to the HR$>-0.2$ $w(\theta)$,
  while the dashed line corresponds to the HR$<-0.2$ $w(\theta)$ fit.}
\label{fig:hr_sa}
\end{figure}

\begin{figure}
\includegraphics[width=0.9\columnwidth]{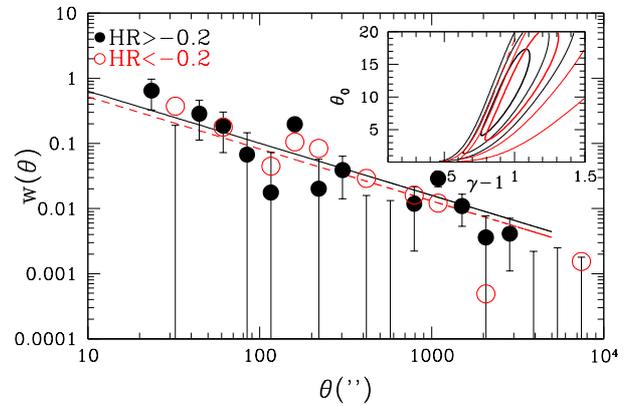}
\caption{As in Figure \ref{fig:hr_sa} but for the hard band.}
%CF for the whole XMM-LSS sample in the hard
%  band for sources with HR$>-0.2$ (Solid, obscured AGN), and for
%  sources with HR$<-0.2$ (lower panel; unobscured AGN).}
\label{fig:hr_ha}
\end{figure}

To this end we compared the correlation function of the 
hard and soft-spectrum sources
by separating them, within each band, using the 
hardness ratio, HR, indicator defined as
\begin{equation}
{\rm HR}=\frac{{\rm CR}_{h}-{\rm CR}_{s}}{{\rm CR}_{h}+{\rm CR}_{s}},
\end{equation}
where CR$_{s}$ and CR$_{h}$ represent the total count rates in the
soft and the hard band, respectively. It is known that most of the
sources with HR$> -0.2$ are likely to be obscured (hard-spectrum) AGN; 
conversely, the sources with HR$< -0.2$ are mostly (soft-spectrum) unobscured 
(see \cite{gandhi04} for details). Using this
criterion, we split the whole sample and derived the
 log $N-$log $S$ distributions for each of them in the soft and the hard
bands  (Fig.~\ref{fig:lnls_hr}). Table~\ref{tab:hr} and
Figs.~\ref{fig:hr_sa}~-~\ref{fig:hr_ha} show the parameters and ACFs for
the obtained subsamples.

\begin{table}[tbh]
\begin{center}
\caption{Correlation function for the subsamples characterized by
  their hardness ratio above and below $-0.2$.}
\label{tab1}
\tabcolsep 2pt
\begin{tabular}{ccccccc}
Band & HR & $N$ & $\theta^{''}_{0}$ & $\gamma $ &
$\theta^{''}_{0,\gamma=1.8}$ & $w(<3.3^{'})$\\
\hline
Soft   & $>$-0.2 & 674  & $10.3\pm 3.3$& $1.93\pm 0.08$ & 5.2$\pm2.0$ & 0.066$\pm 0.048$\\
       & $<$-0.2 & 4418 & $1.5\pm 0.2$ & $1.80\pm 0.02$ & 1.5$\pm0.2$ & 0.019$\pm 0.005$\\
\hline
Hard   & $>$-0.2 & 1170 & $10.7\pm 1.7$ & $1.94\pm 0.04$ & 5.5$\pm 1.0$ & 0.129$\pm0.028$\\
       & $<$-0.2 & 1198 & $13.1\pm 2.4$ & $2.04\pm 0.06$ & 4.4$\pm 1.1$ & 0.056$\pm 0.026$\\
\hline
\label{tab:hr}
\end{tabular}
\end{center}
\end{table}

The main result of this analysis is that there is a
distinct clustering difference between the sources 
with hard and soft spectra in the soft band, with the former sources being significantly more
clustered. In the hard band the corresponding comparison shows a much
weaker difference, in the same direction, but not that
significant. However, one also observes that the integrated signal
within separations $\mincir 3$ arcmin indicates that at least on these small
scales the hard-spectrum sources show a stronger clustering
signal than the corresponding soft-spectrum ones. 
 Similar results were found in \citet{gandhi06}. 
Therefore, one may conclude that indeed there are
indications for a different clustering pattern between hard-spectrum and
soft-spectrum sources, which cannot be attributed to their different
flux-limits, since we verified that this result is valid for
brighter flux-limits as well. We believe that this result suggests a possible
environmental component in the determination of the
different types of AGN, 
beyond their orientation with respect to the observer's
line-of-sight. An environmental dependence of the AGN type has also
been found in local optical AGN samples \citep[e.g.,][ and references therein]{Koul06a,
  Koul06b, Koulouridis}.

\section{Inverting from angular to spatial clustering}
We now derive the spatial correlation length that corresponds
to the measured angular clustering. To this end we used the usual
Limber inversion \citep{peebles}. The main steps are
sketched below.

In a spatially flat universe, the ACF
$w(\theta)$ can be obtained from the spatial one, $\xi(r)$, by
\begin{equation}
w(\theta)=2\frac{\int_{0}^{\infty} \int_{0}^{\infty} x^{4} 
\phi^{2}(x) \xi(r,z) {\rm d}x {\rm d}u}
{[\int_{0}^{\infty} x^{2} \phi(x){\rm d}x]^{2}} \;\; , 
\end{equation}
where the physical separation between any two sources that are 
separated by an angle $\theta$ and considering 
the small angle approximation, is given by
\begin{equation}
\label{eq:approx}
r \simeq \frac{1}{(1+z)} \left( u^{2}+x^{2}\theta^{2} \right)^{1/2} \;,
\end{equation}
while $\phi(x)$ is the selection function (the probability 
that a source at a distance $x$ is detected in the survey) 
given by 
\begin{equation}
\phi(x)=\int_{L_{\rm min}(z)}^{\infty} \Phi(L_{x},z) {\rm d}L \;\;,
\end{equation}
where $\Phi(L_{x},z)$ is the redshift-dependent luminosity function
of the X-ray selected AGN. A variety of X-ray source luminosity
functions are available in the literature, and to investigate
the uncertainty that their differences can introduce in the derived
value of $r_{0}$, we will present results for a number of $\Phi(L_{x},z)$.
Although the most recent soft/hard band luminosity
functions are those of \citet{ebrero}, we will also use those of \citet{Hasinger} for the 
soft band, while for the hard band we used those of \citet{Ueda} and of
\citet{LaFranca2005}. 
In all cases we used of course the luminosity-dependent density
evolution model of the luminosity function.

The proper distance $x(z)$ is related to the redshift through 
\begin{equation}
x(z)=\frac{c}{H_{0}} \int_{0}^{z} \frac{{\rm d}y}{E(y)}\;\; ,
\end{equation}
with 
\begin{equation}
E(z)=[\Omega_{\rm m}(1+z)^{3}+\Omega_{\Lambda}]^{1/2} \;\;\;
\Omega_{\rm m}=1-\Omega_{\Lambda}.
\end{equation}

In this context, the spatial correlation function can 
be modeled as in \citet{dezotti}
\begin{equation}\label{eq:dezotti}
\xi(r,z)=(r/r_{0})^{-\gamma}\times (1+z)^{-(3+\epsilon)}\;\;,
\end{equation} 
where $r_{0}$ is the correlation length in three dimensions and 
$\epsilon (\equiv\gamma-3)$ parameterizes the type of clustering evolution.
A value of $\epsilon=-1.2$ for $\gamma=1.8$, indicates a
constant clustering in comoving coordinates,
%which means that the amplitude of the correlation function
%remains fixed with redshift in commoving coordinates as the galaxy pair 
%expands together with the Universal expansion.
%On the other hand, in the 
while $\epsilon=-3$ indicates a constant clustering in physical coordinates 
%and finally a value of $\epsilon=0$ reflects the stable clustering
%model
\citep[e.g.,][]{dezotti}.

Combining the above system of 
equations, we obtained the following integral equation
for $w(\theta)$ 
\begin{equation}\label{eq:angu}
w(\theta)=2\frac{H_{0}}{c} \int_{0}^{\infty} 
\left(\frac{1}{N}\frac{{\rm d}N}{{\rm d}z} \right)^{2}E(z){\rm d}z 
\int_{0}^{\infty} \xi(r,z) {\rm d}u,
\end{equation} 
where ${\rm d}N/{\rm d}z$ denotes
the number of objects in the given 
survey within a solid angle $\Omega_{s}$ and in 
the shell $(z,z+{\rm d}z)$. It takes the following form:
\begin{equation}
\frac{{\rm d}N}{{\rm d}z}=\Omega_{s}
x^{2}\phi(x)\left(\frac{c}{H_{0}}\right) E^{-1}(z)\;\;.
\end{equation}

Using 
Eq.(\ref{eq:dezotti}), Eq.(\ref{eq:approx}) and Eq.(\ref{eq:angu}), 
we find that the amplitude $\theta_{0}$ in two 
dimensions is related to the 
correlation length $r_{0}$ in 
three dimensions through the equation \citep[see][]{basilakos_2005}:
 
\begin{equation}
\theta_{0}^{\gamma-1}=H_{\gamma}r_{0}^{\gamma}
\left(\frac{H_0}{c}\right) 
\int_{0}^{\infty} \left( \frac{1}{N}\frac{{\rm d}N}{{\rm d}z}
\right)^{2} \frac{E(z)}{x^{\gamma-1}(z)}  (1+z)^{-3-\epsilon+\gamma}
{\rm d}z \;,
\end{equation}
where $H_{\gamma}=\Gamma(\frac{1}{2})
\Gamma(\frac{\gamma-1}{2})/\Gamma
(\frac{\gamma}{2})$.  

\begin{figure}
\includegraphics[width=0.9\columnwidth]{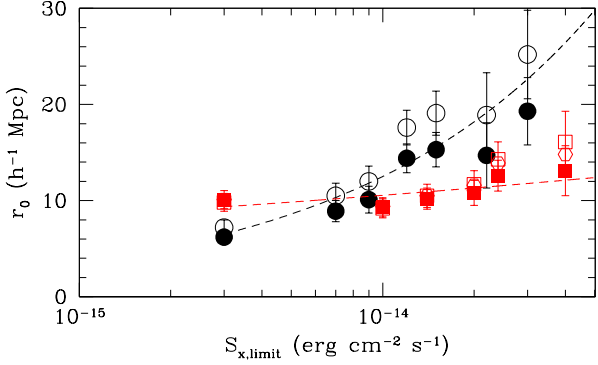}
\caption{Spatial correlation length $r_{0}$ for $\gamma =1.8$ and
  $\epsilon=-1.2$ considering the homogeneous 10 ks
  based observations as a function of the flux limit of the sample in the soft band
(filled and empty circles correspond to the Ebrero and Hasinger
$\Phi_x(L)$, respectively) and in the hard band (filled square, open
squares and open hexagons correspond to the Ebrero, Ueda and La Franca
$\Phi_x(L)$, respectively).}
\label{fig:ro}
\end{figure}

Following the previous steps, we derived the spatial clustering length
scale for fixed $\gamma=1.8$ and for both values of
clustering evolution parameter ($\epsilon=-1.2$ and $-3$). The results
are presented in Table \ref{tab:r0}. Evidently that all three
hard band luminosity functions provide the same $r_0$ value, while
for $\epsilon=-1.2$ there is a difference in the 
soft band with the \citet{Hasinger}
$\Phi_x(L)$, providing an $r_0$ value that is 16\% higher than that
provided by \citet{ebrero} $\Phi_{x}(L)$. As we will see, this
difference increases proportionally to the flux-limit of the subsample used.
\begin{table}[tbh]
\caption{Spatial correlation length $r_{0}$ (in $h^{-1}$ Mpc), provided by Limber's inversion of the
  ACF and using different AGN X-ray luminosity functions, for the
  homogeneous $10~$ks sample and for the lowest flux-limit available. Note
  that the corresponding soft and hard band median redshifts are
  $\bar{z}\simeq 1.1$ and $\simeq 1$, respectively, while the peaks of
  the corresponding redshift distributions are at $z\simeq 1$ and 0.7, respectively.}
\label{tab:r0}
\tabcolsep 5pt
\begin{tabular}{ccc|ccc}
 &\multicolumn{2}{c|}{Soft band} &\multicolumn{3}{c}{Hard band} \\ \hline
$\epsilon$     & Ebrero  &  Hasinger & Ebrero & La Franca & Ueda\\
$-1.2$ &6.2$\pm0.7$ & 7.2$\pm0.8$ &10.1$\pm0.9$ &9.8$\pm0.9$ &10.1$\pm0.9$ \\ 
$-3$   &3.2$\pm0.4$ & 3.3$\pm0.4$& 5.3$\pm0.5$& 5.2$\pm0.5$& 5.3$\pm 0.5$\\ \hline
\end{tabular}
\end{table}
In Fig. \ref{fig:ro} we present the inverted $r_0$ values as a
function of the different flux limits, as they appear in Fig. \ref{fig:limit}. 
We see that for the soft band the two luminosity functions used 
in the inversion provide $r_0$ values that diverge with increasing flux-limit. 

The dashed lines in Fig. \ref{fig:ro} correspond to fits of the data, using for each band results based on all different
luminosity functions, of the form:
\begin{equation}
r_0=A \left(\frac{f_x}{3 \times 10^{-15}}\right)^\beta,
\end{equation}
with $(A, \beta)\simeq (6.5, 0.54)$ for the soft band and 
$(A, \beta)\simeq (9.4, 0.1)$ for the hard band.
Evidently, the flux dependence of clustering, once one inverts from angular to 3D
space, is preserved mostly in the soft band. In the hard band
we see at most a weak dependence  and only for fluxes $\magcir 2
\times 10^{-14}$ erg s$^{-1}$ cm$^{-2}$, while a constant hard band value of
$r_0 \simeq 10 \;h^{-1}$ Mpc, irrespective of the flux limit, 
 appears also to be consistent with the data. These hard band results
 agree with those of \citet{ebrero}, who found that the
 weak dependence of $\theta_0$ on the flux-limit translates into a roughly
 constant $r_0$ as a function of flux-limit, or equivalently as a
 function of median redshift or median X-ray luminosity of the
 sample. However, a relatively strong dependence of the soft band $r_0$ with respect
 to the flux-limit disagree with  \citet{ebrero},
but agrees with \citet{Plionis08}.

\section{Bias of the X-ray selected AGN}
The concept of biasing between different classes of extragalactic objects 
and the background matter distribution was introduced by \citet{Kaiser}
and \citet{Bardeen} to explain the higher amplitude of the two-point correlation function 
of clusters of galaxies with respect to that of galaxies themselves.
In our case and within the framework of
linear biasing \citep[cf.][]{Kaiser, Benson}, 
the evolution of the bias parameter is usually defined as
\begin{equation}
\label{eq:biasdef}
b^2(z)=\frac{\xi_{AGN}(8,z)}{\xi_{DM}(8,z)}=\left[\frac{r_{0}(z)}{8}\right]^{\gamma}
\frac{1}{\xi_{DM}(8,z)},
\end{equation}
where $\xi_{AGN}(8,z)=(r_{0}(z)/8)^{\gamma}$ and $\xi_{DM}(8,z)$ are 
the spatial correlation functions of AGN and dark matter halos 
evaluated at 8 $h^{-1}$~Mpc, respectively. 
Notice that the correlation lengths in 3D 
%and the median redshifts values ($z=\hat{z}$) 
are presented in Table \ref{tab:r0}. The correlation function of the DM halos
is given by \citet{peebles}
\begin{equation}
\label{eq:sigma8}
\xi_{DM}(8,z)=\frac{\sigma_{8}^{2}(z)}{J_2},
\end{equation}
where $J_2=72/\left[(3-\gamma)(4-\gamma)(6-\gamma)2^\gamma\right]$ 
and $\sigma_8^2(z)$ is the dark matter density variance in a 
sphere with a comoving radius of 8 $h^{-1}$~Mpc, which evolves as
\begin{equation}
\label{eq:sigma8z}
\sigma_{8}(z)=\sigma_{8}D(z)/D(0) \;\;.
\end{equation}
Note that $D(z)$ is the linear growth factor scaled to unity
at the present time. For the concordance   
$\Lambda$ cosmology\footnote{In this work we use 
$\Omega_{\rm m}=1-\Omega_{\Lambda}=0.3$ and $\sigma_{8}=0.80$.} 
the growth factor becomes  \citep[see][]{Peebles1993}
\begin{equation}
\label{eq24}
D(z)=\frac{5\Omega_{\rm m}
  E(z)}{2}\int^{+\infty}_{z}
\frac{(1+y)}{E^{3}(y)}\;dy \;\;. 
\end{equation}
Finally, inserting 
Eq.(\ref{eq:sigma8z}) and Eq.(\ref{eq:sigma8}) 
into Eq.(\ref{eq:biasdef}), we obtain
the evolution of biasing with epoch as a function of the
clustering properties
\begin{equation}
\label{obsbias24}
b(z)=\left[ \frac{r_{0}(z)}{8}\right]^{\gamma/2}
\frac{J_{2}^{1/2}}{\sigma_{8} D(z)/D(0)} \;.
\end{equation}

For angular clustering we may identify the dominant
redshift of the sample under study as that predicted by the luminosity
function of the sources used and the flux limit of the sample, which
predicts the redshift distribution of the sources. We can then obtain
from the last equation an estimate of the bias of our X-ray
sources (see Table \ref{tab:bias}). Of course one has to keep in mind
that this is a quite crude estimate since we implicitly assume
that all detected sources obey the same luminosity function, while
in effect luminosity functions are derived from subsamples of all 
the detected X-ray sources for which optical counterparts are identified.

\begin{table}[tbh]
\caption{Linear bias factor for the lowest flux-limit results of the
  homogeneous $10~$ks XMM-LSS data (and for the same X-ray luminosity
  functions as in Table 6).}
\label{tab:bias}
\tabcolsep 5pt
\begin{tabular}{ccc|ccc}
 &\multicolumn{2}{c|}{Soft band} &\multicolumn{3}{c}{Hard band} \\ \hline
$\epsilon$     & Ebrero  &  Hasinger & Ebrero & La Franca & Ueda \\
$-1.2$ &2.2$\pm0.2$ & 2.7$\pm0.3$ &3.3$\pm0.3$ &3.2$\pm0.3$ &3.3$\pm0.3$ \\ 
$-3$   &1.2$\pm0.1$ & 1.3$\pm0.1$& 1.9$\pm0.2$& 1.8$\pm0.2$& 1.9$\pm 0.2$\\ \hline
\end{tabular}
\end{table}
We see again that although our hard band results roughly agree with
those of \citet{ebrero}, our soft band results are significantly
different, because we found a significantly
weaker clustering amplitude than the aforementioned authors. 

We can now use a bias evolution model \citep[e.g.,][and references therein]{Sheth, BPR}
to estimate the halo mass that corresponds to the above
estimated bias factors (for $\epsilon=-1.2$), assuming that each halo hosts one AGN
source. Using the latter model (see details in Papageorgiou, Plionis,
Basilakos \& Ragone-Figueroa {\em in prep.}), we obtain that for the soft band
and the Ebrero et al. luminosity function the corresponding halo mass
is $M_h\simeq 10^{12.9\pm 0.3} \; h^{-1} M_{\odot}$, while using the Hasinger
luminosity function the corresponding value is $M_h \simeq 10^{13.2\pm 0.3}
\; h^{-1} M_{\odot}$. For the hard band we find that $M_h\simeq
10^{13.7\pm 0.3} \; h^{-1} M_{\odot}$. Note that using the Sheth et al. bias model, we find very
similar $M_h$ values (for example, for the hard band results we
find $M_h\simeq 10^{13.6} \; h^{-1} M_{\odot}$).

\section{Main conclusions}

We have performed a two-point correlation function analysis
of the XMM-LSS sample of point sources %and Subaru fields 
that contains in total 94 XMM-Newton
pointings (more than five thousand point-like sources). 
The observations were made near the celestial equator at high galactic
latitudes over $\sim 11$ sq. deg. in the soft (0.5-2 keV) and hard
(2-10 keV) bands with effective exposures ranging from 8.1 to 47.3 ks. The
minimum flux limits are almost $10^{-15}$ and $3\times 10^{-15}$ erg
s$^{-1}$ cm$^{-2}$ for the soft and hard bands, respectively. For the
definition of the detection probabilities for each source and for the
proper generation of the mock catalogs we performed a series of
numerical Monte-Carlo simulations of the XMM-Newton observations.
The most important points and results of our work are listed below.

To deal with the pointing overlap question, we considered two approaches: that of a $10'$ off-axis limitation,
 and the Voronoi delimitation. No major differences were observed in the derived
point-source correlation function between these
two approaches. We consequently followed the statistically richer
Voronoi delimitation approach, which produces a contiguous field.

The $\log N-\log S$ distributions for the soft and hard bands were
found to agree well with the results from the previously released XMM-LSS catalog
\citep{gandhi06}. Using the whole exposure XMM-LSS data, 
we extended the $\log N-\log S$ to lower fluxes, ie., 
$10^{-15}$ and $3\times 10^{-15}$ erg s$^{-1}$ cm$^{-2}$ for the soft and
hard bands, respectively.

The amplitude of the correlation function $w(\theta)$ is
  significantly higher in the hard band than in the soft band at the
  lowest fluxes. When analyzing a homogeneous 10 ks extracted sample
  from the full exposure data, this difference becomes more
  prominent. At higher fluxes ($f_x\magcir 10^{-14}$ erg s$^{-1}$
  cm$^{-2}$) the amplitude of the correlation function becomes higher
  in the soft band. These results provide a bias
  factor at a median redshift $\bar{z}\simeq 1.1$ of $\sim 2.5$ for
  the soft band when inverted to 3D (and for
  $\epsilon=-1.2$), 
  and at $\bar{z}\simeq 1$ of $\sim 3.3$ for the hard band sources. 
  These bias values correspond to
  a mass of the halos hosting the AGN sources of $M_h\sim 10^{13\pm 0.3}
\; h^{-1} M_{\odot}$ for the soft band and $M_h\sim 10^{13.7\pm 0.3}\; h^{-1}
  M_{\odot}$ for the hard band.

The correlation at degree-scale ($\magcir 3000''$)  nicely
    extends that observed on an arcmin scale ($100-1000''$), 
   a result which is obtained thanks to the wide contiguous
    area covered by the survey.

The hard-spectrum sources show a stronger clustering than the
soft-spectrum ones, especially in the soft band. This hints at
an environmental dependence of the AGN type.

The amplitude of the spatial correlation function 
  grows with flux limit, but mostly in the soft band. In the hard band
  there is at most a weak dependence, with a constant value of $r_0 \simeq 10
  \;h^{-1}$ Mpc, which is consistent with the data.

\begin{acknowledgements}
The simulations were performed at the CNRS "Centre de Calcul de l'IN2P3" located in 
Lyon, France. The authors would like to thank Pierrick Micout for his
help regarding the use of the CC-IN2P3. AE, OM, EG and JS acknowledge
support from the ESA PRODEX Programme "XMM-LSS", from the "Belgian
Federal Science Policy Office" and from the "Communaut\'e fran\c{c}aise de
Belgique - Actions de recherche concert\'ees -
Acad\'emie universitaire Wallonie-Europe".

\end{acknowledgements}

\end{document}